\documentclass[acmsmall,printacmref=false,printfolios=false,nonacm=true]{acmart}

\usepackage{bbm}
\usepackage{graphicx}
\usepackage{subcaption}
\definecolor{celadon}{rgb}{0.67, 0.88, 0.69}
\definecolor{darkseagreen}{rgb}{0.56, 0.74, 0.56}
\usepackage{pgfplots}\pgfplotsset{compat=1.9}
\usepackage{wrapfig}
\usepackage{subcaption}
\usepackage{multirow}
\usepackage{multicol}
\usepackage{color}
\usepackage{tikz}
\usetikzlibrary{babel}
\usetikzlibrary{trees}
\usetikzlibrary{external}
\AtBeginDocument{%
  \providecommand\BibTeX{{%
    \normalfont B\kern-0.5em{\scshape i\kern-0.25em b}\kern-0.8em\TeX}}}

\begin{document}

\title{Fairness in Recommendation: Foundations, Methods and Applications}

\author{Yunqi Li}
\affiliation{%
  \institution{Rutgers University}
  \country{New Brunswick, NJ, US}}
\email{yunqi.li@rutgers.edu}

\author{Hanxiong Chen}
\affiliation{%
  \institution{Rutgers University}
  \country{New Brunswick, NJ, US}}
\email{hanxiong.chen@rutgers.edu}

\author{Shuyuan Xu}
\affiliation{%
  \institution{Rutgers University}
  \country{New Brunswick, NJ, US}}
\email{shuyuan.xu@rutgers.edu}

\author{Yingqiang Ge}
\affiliation{%
  \institution{Rutgers University}
  \country{New Brunswick, NJ, US}}
\email{yingqiang.ge@rutgers.edu}

\author{Juntao Tan}
\affiliation{%
  \institution{Rutgers University}
  \country{New Brunswick, NJ, US}}
\email{juntao.tan@rutgers.edu}

\author{Shuchang Liu}
\affiliation{%
  \institution{Rutgers University}
  \country{New Brunswick, NJ, US}}
\email{shuchang.liu@rutgers.edu}

\author{Yongfeng Zhang}
\affiliation{%
  \institution{Rutgers University}
  \country{New Brunswick, NJ, US}
}
\email{yongfeng.zhang@rutgers.edu}

\renewcommand{\shortauthors}{Y. Li, H. Chen, S. Xu, Y. Ge, J. Tan, S. Liu and Y. Zhang}

\begin{abstract}
As one of the most pervasive applications of machine learning, recommender systems are playing an important role on assisting human decision making. The satisfaction of users and the interests of platforms are closely related to the quality of the generated recommendation results. However, as a highly data-driven system, recommender system could be affected by data or algorithmic bias and thus generate unfair results, which could weaken the reliance of the systems. As a result, it is crucial to address the potential unfairness problems in recommendation settings.
Recently, there has been growing attention on fairness considerations in recommender systems with more and more literature on approaches to promote fairness in recommendation. However, the studies are rather fragmented and lack a systematic organization, thus making it difficult to penetrate for new researchers to the domain. This motivates us to provide a systematic survey of existing works on fairness in recommendation. This survey focuses on the foundations for fairness in recommendation literature. It first presents a brief introduction about fairness in basic machine learning tasks such as classification and ranking in order to provide a general overview of fairness research, as well as introduce the more complex situations and challenges that need to be considered when studying fairness in recommender systems. After that, the survey will introduce fairness in recommendation with a focus on the taxonomies of current fairness definitions, the typical techniques for improving fairness, as well as the datasets for fairness studies in recommendation. The survey also talks about the challenges and opportunities in fairness research with the hope of promoting the fair recommendation research area and beyond.
\end{abstract}



\maketitle


\section{Introduction}
Recommender systems are playing an essential role in the era of information explosion. They help users to gain access to the information they are interested in and benefit sellers or producers to increase their exposure so as to gain profits. Since recommender systems are associated with multi-sided benefits by deciding what information to be delivered, fairness becomes an especially critical problem that requires attention. However, in the literature on fairness research, researchers have shown that the recommender system, as a highly data-driven application, can suffer from various unfairness issues and may harm the benefits of multiple stakeholders
\cite{xiao2017fairness,yao2017beyond, burke2017multisided, abdollahpouri2019multi, leonhardt2018user, beutel2019fairness,li2021user,ge2021towards,li2021counterfactual}. For example, an unfair job recommender system may exhibit racial or gender discrimination by disproportionately recommending low-payment jobs to certain (protected) user groups; e-commerce recommender systems may disproportionately recommend the products from popular sellers while providing very limited exposure opportunity to other equally qualified but less popular items; news recommender systems may over recommend specific political ideologies over others (sometimes due to echo chambers) which may manipulate user's opinions. To improve the satisfaction of various participants, it is important to solve the unfairness issue in recommender systems to build a positive and sustainable ecosystem. In recent years, 
works on fair recommendation have come to the fore and received increasing attention. The goal of this survey is to synthesize the current state of fairness in recommendation and to inspire more future work in this area.

We begin the survey with a brief introduction of fairness in machine learning to provide the readers a general and basic background knowledge of fairness research. We talk about the main causes of unfairness, the typical methods for achieving fairness, the diverse fairness considerations in machine learning and their relationship with the theories of justice in ethics. Though focusing on fairness in recommendation, we first talk about certain fairness concerns and representative methods in classification and ranking tasks due to the following considerations: first, we would like to take fairness in basic machine learning tasks such as classification and ranking as examples to help readers better understand the key concepts introduced in the previous research; second, before studying fairness in recommendation, the first endeavor to achieve fairness in the community is to develop fair classifiers and rankers. Therefore, these two areas provide insightful knowledge and can be enlightening for the researches in fair recommendation since recommendation problems can usually be formulated as classification or ranking tasks.

     
\begin{wrapfigure}{r}{0.3\textwidth}
\vspace{-2ex}
\resizebox{0.3\textwidth}{!}{
\includegraphics[scale=0.5]{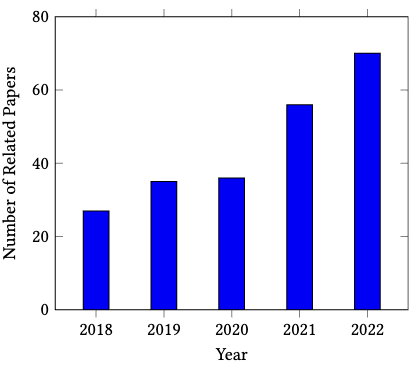}
}
\vspace{-5ex}
\caption{The number of related papers in the last five years.}
\label{fig0}
\end{wrapfigure}

However, promoting fairness in recommendation can face unique challenges. For example, recommender systems often consist of multiple models to balance multiple goals; recommender systems are dynamic and need to consider long-term benefits; the extreme data sparsity can also bring additional difficulties for model learning and evaluating. What's more, unlike fairness in classification and ranking problems that usually consider a single-side fairness requirement, the concept of fairness in the recommendation has been extended to multiple-stakeholders \cite{burke2017multisided} and consequently make the fairness problem more challenging. In the main body of the survey, we introduce fairness in recommendation from four perspectives: \textit{taxonomy}, \textit{techniques}, \textit{datasets} and \textit{open challenges}. As aforementioned, recommender systems consider complicated application scenarios so that researchers view unfairness issues from very different perspectives. This motivates us to provide a systematical taxonomy of fairness notions in recommendation to help readers have a clear understanding of how to consider unfairness problems in recommendation scenario. After that, we introduce the typical techniques for promoting fairness in recommendation to help explain how to alleviate the various unfairness concerns. Additionally, we collect and present several publicly available datasets with user or item sensitive features to benefit future fairness studies in recommendation. Last but not the least, we summarize open challenges and opportunities in fair recommendation research to suggest future directions of this area. 




In recent years, there has been an increasing number of articles on fair recommendation as we showed in Figure.\ref{fig0}. The scope of this survey covers more than 260 works including the representative papers about fairness studies in AI and the papers about fairness in recommendation published in top Web, IR, RecSys, and Data Mining related conferences and journals such as WWW, SIGIR, KDD, RecSys, CIKM, WSDM, AAAI, IJCAI, FAccT, FnTIR, TOIS, TORS, TKDE, TIST, just to name a few, as well as some of the outstanding arXiv papers.

The remaining of this survey is organized as follows. We first introduce some related surveys and demonstrate the difference and contributions of this work in Section \ref{related}. We provide a general introduction about fairness in machine learning in Section \ref{ml}. We introduce the theories of justice and discuss its relationship with fairness notions in machine learning in Section \ref{Justice}. Section \ref{classification} and Section \ref{ranking} briefly talk about fairness studies in classification and ranking tasks, respectively. Section \ref{rec} discusses fairness in recommendation from various perspectives under a comprehensive taxonomy. We discuss some important challenges and opportunities of fairness research in recommendation in Section \ref{challenges}, and Section \ref{con} concludes the survey.









\section{Related Surveys}\label{related}
In the past few years, a number of surveys talking about fairness and bias in general machine learning have been published  \cite{selbst2019fairness, caton2020fairness, mehrabi2021survey, wan2021modeling, pagano2022bias, castelnovo2022clarification}. However, they usually focus on the fairness works in classification tasks. Some other surveys pay attention to one certain field in machine learning or fairness works, for example, \citeauthor{shi2021survey} \cite{shi2021survey} provide a survey to help readers gain insights into fairness-aware federated learning, \citeauthor{zhang2021fairness} \cite{zhang2021fairness} focus on the literature of fairness in data-driven sequential decision-making, and \citeauthor{makhlouf2020survey} \cite{makhlouf2020survey} review a list of causal-based fairness notions and study their applicability in real-world scenarios. A few surveys also provide an overview of fairness in ranking tasks \cite{zehlike2021fairness, patro2022fair}. Recommendation algorithms can usually be considered as a type of ranking algorithm. However, existing works on fair ranking usually only consider unfairness issue from the item perspective, such as the item exposure fairness, while the fairness concept in recommender systems can be more complicated and needs to be considered from multiple sides \cite{burke2017multisided}. \citeauthor{pitoura2021fairness}~\cite{pitoura2021fairness} talks about fairness in both ranking and recommendation, and \citeauthor{ekstrand2022fairness} \cite{ekstrand2022fairness} discusses fairness in information access systems such as information retrieval and recommendation. Though covering a brief introduction about fairness in classification and ranking, our survey pays specific attention to organizing the concept of fairness in recommendation through a comprehensive taxonomy of fairness notions proposed in recommendation problems, the task-specific techniques for promoting recommendation fairness, as well as the datasets specially suitable for fairness research in recommendation. Moreover, \citeauthor{chen2020bias} \cite{chen2020bias} provides a survey on bias and debias in recommender system, which covers a part of content about fairness in recommendation. However, most of the works on bias in recommendation focus on improving the recommendation accuracy or robustness in out-of-distribution scenario through debiasing methods instead of promoting fairness. Some surveys such as \cite{ge2022survey, fan2022comprehensive} discuss the trustworthy perspectives of recommender systems which includes fairness as a sub-category of trustworthiness. Our survey differs from these works as we provide a broader and more in-depth introduction of fairness in recommendation. We introduce both technical and philosophical foundations of fairness, better situate recommendation fairness under machine learning fairness context, and introduce a comprehensive taxonomy of recommendation fairness research.

Recently, two surveys on fairness in recommendation were released concurrently as ours \cite{deldjoo2022survey, wang2022survey}. 
Compared to these works, our survey offers the following benefits: 1) We better situate recommendation fairness under the context of machine learning by providing a high-level overview of fairness studies in machine learning, the intuition behind the fairness notions in machine learning, and the timeline of how fairness research develops in the AI field from classification, ranking, to recommendation tasks, so that readers who are new to the fairness in recommendation field can effectively build a systematic structure of knowledge without having to read other materials in advance; 2) We not only introduce the technical approaches to fairness in recommendation, but also introduce the philosophical foundations of fairness research by introducing how it relates to the justice theories in philosophical ethics; 3) We offer a thorough taxonomy for fairness in recommendation research, which helps readers to understand the different fairness considerations in recommender systems and to build a better organized structure of the literature in the field.

\begin{figure}
    \centering
    \includegraphics[scale=0.5]{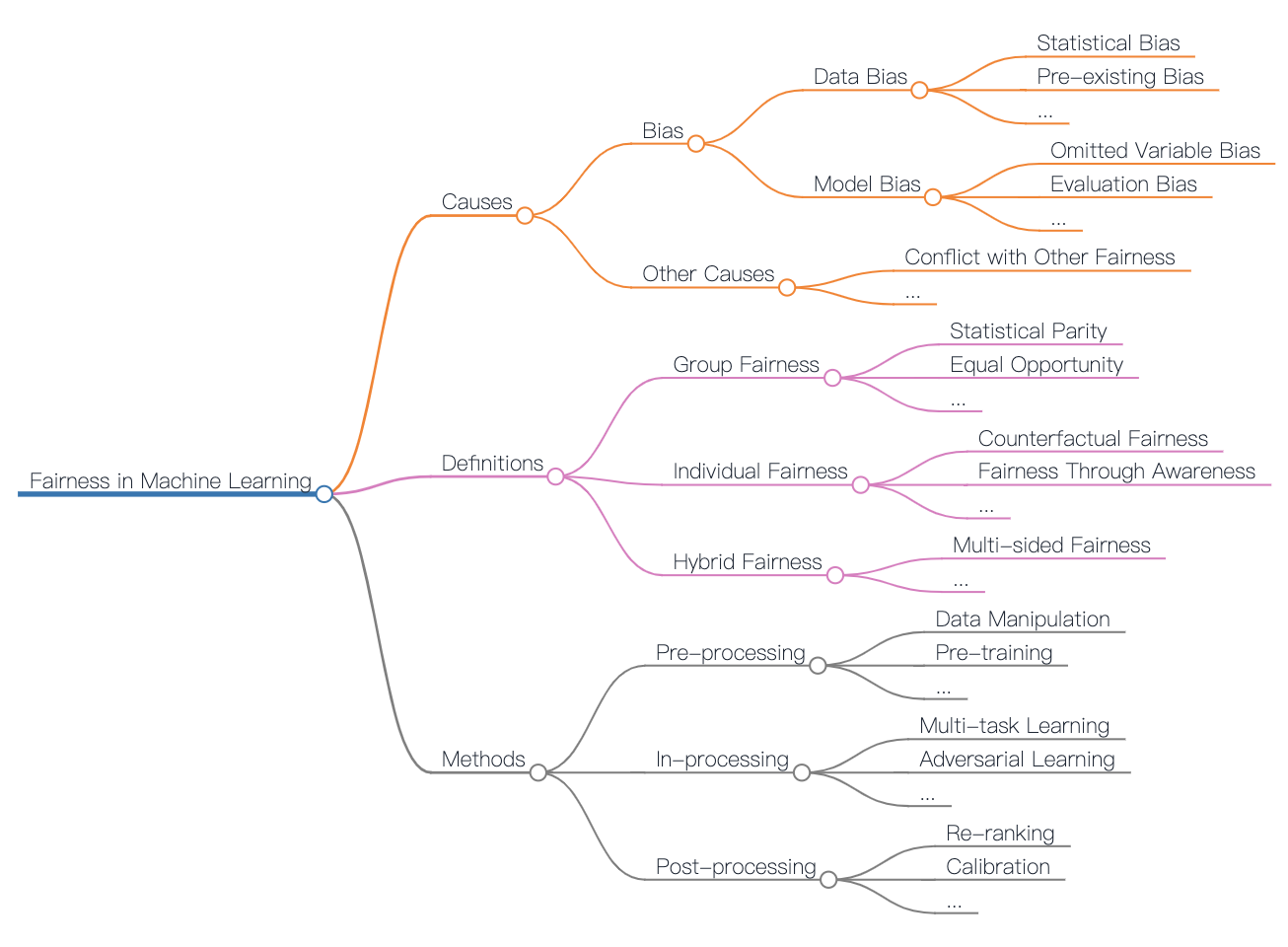}
    \vspace{-15pt}
    \caption{A Basic Structure of Fairness in Machine Learning and Recommender Systems}
    \label{mlbasic}
\end{figure}

\section{Fairness in Machine Learning}\label{ml}
In this section, we provide basic background knowledge of fairness in machine learning, including the causes of unfairness, diverse fairness definitions, as well as the fair learning methods. A mind map of fairness in machine learning and recommender systems is shown in Figure.\ref{mlbasic}.


\subsection{The Causes of Unfairness}

Unfairness in machine learning is usually caused by various forms of biases. A list of different sources of biases with their corresponding definitions in machine learning has been provided in several previous surveys \cite{olteanu2019social, suresh2019framework, mehrabi2021survey, castelnovo2022clarification}. It is out of the scope of this survey to list and discuss all of the types of biases in machine learning, nevertheless, we want to talk about the main sources of biases from a higher level. Generally speaking, as the two fundamental components of machine learning system, training data and learning process are also the main sources of biases which can lead to unfair results in machine learning tasks. Therefore, among all kinds of biases, we categorize them into two types: \textit{Data Bias} and \textit{Model Bias}.
In specific, \textit{Data Bias} covers the types of biases lying in the training data itself, such as the biases coming from the processes of data generation, data collection, and data storage; meanwhile, \textit{Model Bias} includes biases which are not present in the input data, but introduced in the processes of model designing, model training, and model evaluation.

\subsubsection{\textbf{Data Bias}}
Most of the biases in machine learning lie in the data itself. The bias in data can come from the processes of data generation, data collection, and data storage, etc. Here we roughly categorize the data bias into \textit{Statistical Bias} and \textit{Pre-existing Bias}.

\textit{Statistical Bias} \cite{castelnovo2022clarification}. The statistical bias usually arises from the process of data collection, storage or cleaning. It occurs when there are flaws in the experimental design or data collection process and thus the data is not the true representation of the population. For example, the data is not randomly selected from the full population, is wrongly recorded, or is systematically missing, resulting in the lack of population diversity or other anomalies. Statistical bias can easily lead to systematic differences between the true parameters and the estimated statistics of a population.

\textit{Pre-existing Bias} \cite{suresh2019framework}. Even if the data is perfectly sampled and selected, the bias can also be pre-existing in the data during the generation process. For example, pre-existing bias can occur when the data itself reflects biased decisions, which usually leads to the system being no longer objective and fair. Suppose a company recruiter tends to hire employees with certain race, and the company wants to train a hiring decision system based on its previous recruitment records. The learned system will have a systematic favour towards certain races of people when making hiring decisions. In this case, the AI system will reinforce the pre-existing bias encoded by human decision makers.

\subsubsection{\textbf{Model Bias}}
Moreover, unfairness can arise from the biases in model designing, model training and model evaluating processes, such as the biased model architecture, improper use of certain optimization methods or estimators, and inappropriate benchmarks \cite{baeza2018bias}. Here we list two representative biases from model designing and model evaluation which may bring unfairness and affect user satisfaction.

\textit{Omitted Variable Bias} \cite{data36, mehrabi2021survey}. The omitted variable bias happens when some important variables or features are not considered when designing and training the model. For example, suppose there is a study that investigates the gender pay gap. If the study only considers variables such as education and work experience, but fails to consider the differences in negotiation skills, it may lead to an omitted variable bias. Omitting important variables that are related to both gender and income can result in biased estimates of the gender pay gap, since the real impact of gender on income may be confounded by those unmeasured factors.


\textit{Evaluation Bias} \cite{suresh2019framework}. The evaluation bias usually occurs when inappropriate benchmarks are used in model evaluation. Examples include the Adience and IJB-A (IARPA Janus Benchmark A) benchmarks, which are biased to skin color and gender when evaluating facial recognition systems \cite{mehrabi2021survey}. 

It is worth noting that the biases are intertwined due to the feedback loop phenomenon \cite{chouldechova2018frontiers}. The data used for training machine learning models are usually collected from user behaviors which may present inherent biases, however, the user behaviors can also be affected by the biased learning algorithms, resulting in the further introduction of bias in the data generation process. Therefore, it is important to consider how the biases affect each other to solve them accordingly \cite{mehrabi2021survey}. 

\subsubsection{\textbf{Other Causes}}
It is important to notice that bias may not be the only reason for unfairness and there could be other causes, as a result, researchers should always take caution. One example is the conflict between different fairness requirements. Researches have shown that some fairness requirements cannot be satisfied simultaneously \cite{chouldechova2017fair, kleinberg2016inherent, pleiss2017fairness}, therefore, the violation of one type of fairness may be caused by ensuring another. We will introduce more details on the relationship between fairness definitions in the following parts of the survey.

\subsection{Fairness Definition}

An accepted fact in fairness study is that there is no consensus on fairness definitions since the fairness demands can be different under different scenarios. The fairness concerns can be put forwarded from various perspectives, and it has been theoretically proven that some fairness requirements cannot be satisfied at the same time \cite{chouldechova2017fair, kleinberg2016inherent, pleiss2017fairness}. Therefore, it is important to carefully choose the fairness measurements according to specific problems and scenarios, and be mindful of the potential trade-off between fairness demand and model accuracy. Up to now, many fairness definitions have been proposed in the literature. In general, the definitions of fairness can be classified into three categories: \textit{Group Fairness}; \textit{Individual Fairness}; and \textit{Hybrid Fairness} \cite{wan2021modeling}. The measurement and evaluation of fairness may vary depending on its specific definition. We first provide a general description of the three types of fairness definitions here, and the specific examples in different machine learning tasks will be introduced in the following sections. 

\subsubsection{\textbf{Group Fairness}}

In the study of group fairness requirements, the research subjects are usually divided into different groups based on a certain grouping method, and the most common way is to split users by their sensitive features. 
There is no once-and-for-all answer regarding which features are considered sensitive features since different people may have different definitions, while the connotation of sensitive features usually refers to the features that human-beings cannot choose freely at birth or throughout their life such as biological sex, race and age. Group fairness requires that the protected groups should be treated similarly as the advantaged groups by the machine learning systems \cite{caton2020fairness, wan2021modeling, mehrabi2021survey, castelnovo2022clarification}. For example, to study the unfairness problem of gender discrimination in a hiring decision-making system, the candidates will be first spilt into different groups according to their gender, and then fairness can be measured through computing the difference of system treatments among different groups such as salary or hiring rate.

\subsubsection{\textbf{Individual Fairness}}

Different from group fairness, individual fairness requires that similar individuals should be treated similarly \cite{caton2020fairness, wan2021modeling, mehrabi2021survey, castelnovo2022clarification}. To guarantee individual fairness, we usually first need to determine whether two individuals are similar according to a certain metric, for example, according to the distance of user representations.
The similarity of individuals could be determined by certain sensitive features or combination of features. For example, we may consider all male users whose ages fall into the same range as similar individuals. Sometimes, we may also use latent features to determine user similarity, e.g., we may first represent each user as a latent representation vector according to their profiles or interaction records, and then decide similar users based on vector similarity.
Still consider the above example which is to study the unfairness problem in a hiring decision-making system. We can first represent each candidate as a vector and then require that those candidates with close representations have similar salaries.

It is worthwhile to note that group fairness and individual fairness are two orthogonal concepts. For example, suppose we put similar individuals into the same group, individual fairness emphasizes that individuals within the same group are treated fairly, while group fairness emphasizes that different groups are treated fairness on aggregated group-level. An individually fair method may not be fair group-wise, since it is possible that people within the same group are treated fairly but some groups are much better than other groups. On the other hand, a group-wise fair method may not be individually fair, since it is possible that the groups are treated fairly in terms of group-aggregated metrics but some individuals in a group are treated much better than others in the same group. As a result, neither group fairness nor individual fairness subsume the other, and they both need to be considered carefully.

\subsubsection{\textbf{Hybrid Fairness}}

Instead of focusing on one certain fairness consideration, hybrid fairness recognizes the fact that the fairness demands vary in most cases and aims to achieve more than one fairness requirements simultaneously. For example, in fair recommendation studies, fairness demands may come from both user-side and item-side, thus multiple fairness definitions should be satisfied at the same time \cite{mehrotra2018towards, abdollahpouri2019multi}. However, it is worth noting that not all fairness definitions can be guaranteed at the same time since they sometimes may essentially contradict with each other \cite{chouldechova2017fair, kleinberg2016inherent, pleiss2017fairness}.

\subsection{Methods for Fair Machine Learning} \label{2.2}
Generally, methods that achieve fairness in machine learning fall under three categories: \textit{Pre-processing}, \textit{In-processing}, and \textit{Post-processing} \cite{d2017conscientious, zafar2019fairness, caton2020fairness, mehrabi2021survey}.

\subsubsection{\textbf{Pre-processing Method}}

As the bias issue is often in data itself, a straightforward way is to pre-process the training data before the learning process to remove the underlying bias. The key idea is to train a model on a ``corrected'' data so that the model can be naturally fair. The pre-processing methods are usually achieved through changing the label values for certain data points, or mapping the data to a transformed space. The advantages of pre-processing methods are that the transformed data can be used to train any downstream algorithms without certain assumptions. However, it may suffer unpredictable loss in accuracy and may not remove unfairness on the test data. What's more, the pre-processing method can only be applied if modifying the training data is allowed, both in technical and legal senses.

\subsubsection{\textbf{In-processing Method}}

The in-processing methods usually try to balance the accuracy and fairness demands through modifying the learning process. Most of the existing works of fairness in machine learning use this way to remove discrimination. The methods are often achieved through incorporating fairness metrics into the objective function of the main learning task. The in-processing methods usually show good performance on both accuracy and fairness and provide higher flexibility to trade-off between them. The main disadvantage is that such strategy usually leads to a non-convex optimization problem and cannot guarantee optimality.

\subsubsection{\textbf{Post-processing Method}}

The post-processing methods apply transformations to the model output so as to mitigate unfairness. This is usually achieved by reassigning the labels assigned by the base models, such as recomputing the scores or re-ranking the output lists. The post-processing methods usually treat the base model as a black-box and provide model-agnostic flexibility. Such methods do not need to modify the model and training data, and can achieve relatively good performance and fairness. However, the post-processing methods cannot be used when sensitive feature information is unavailable at the decision time (test-time) since they usually need to access the sensitive information for post-processing.

In short, there are various ways to improve fairness in machine learning, and each method has its own pros and cons. Choosing which method(s) to use is sometimes not a pure technical problem, but the social and legal context must also be considered. In the following sections, we will introduce examples of using pre-processing, in-processing, and post-processing methods for improving fairness.

\section{Philosophical Foundations of Fairness in Machine Learning: The Theories of Justice}
\label{Justice}

Nowadays, we have seen a growing number of works talking about enhancing fairness in machine learning and AI, but few works talk about the philosophical foundations of fairness considerations in machine learning. 
In this section, we will briefly introduce the concept of justice and its relationship with fairness in machine learning so as to help readers understand the philosophical foundation of fairness research in machine learning and recommender systems.

It is noteworthy that justice takes on different meanings in different practical contexts. Therefore, it is crucial to enquire about the fundamental ideas of justice that underlie all of the different applications \cite{miller2021}. In this survey, we follow \cite{miller2021, tang2022and} to introduce the basic theories of justice. Moreover, the understanding about the relationship between justice and fairness also varies in different articles. Currently, the terms "justice" and "fairness" are frequently used interchangeably because of their close relationship, however, some works argues that justice and fairness are not the same thing, for example, \citeauthor{goldman2015justice} \cite{goldman2015justice} demonstrate that \textit{“justice should refer to whether one adheres to certain rules or standards, while fairness should refer to how one responds to perceptions of these rules and rule compliance."} In this survey, we would not debate too much about how justice and fairness differ in non-AI fields; instead, we will concentrate on how the justice concept in ethics relates to fairness studies in machine learning. Basically, we will see that the intuitions behind the concepts of algorithmic fairness in machine learning is often inspired by the principle of justice \cite{miller2021, tang2022and}. Justice as the philosophical origin of fairness in machine learning is firstly discussed in \cite{tang2022and}, here we convey similar ideas and extend the discussion about fairness in machine learning and recommendation research.

\subsection{The Theories of Justice}
\citeauthor{miller2021} \cite{miller2021} summarizes three overarching theories of justice which can help to unify the various types of justice: \textit{Utilitarianism}, \textit{Contractarianism}, and \textit{Egalitarianism}. The utilitarianism understanding of justice lies in supplying the greatest happiness principle to guide actions. Utilitarianism seeks to maximize overall utilities for the greatest number of people while ignoring the concern that how the utilities are allocated among individuals \cite{miller2021}. The contractarianism view of justice focuses on building hypothetical principles and agreements that people would choose to adopt. Making such agreements in real world is challenging since people are likely to express the principles they would prefer from quite radical standpoints \cite{miller2021}. Another theoretical framework to understand justice is egalitarianism, which requires a substantively equal distribution of advantages. The main challenge for the egalitarianism theory is to clarify what kind of equality that justice requires in varying circumstances \cite{miller2021}. Broadly speaking, most of the fairness notions put forth in machine learning literature are consistent with the egalitarianism theory \cite{tang2022and}, for example, the Equal Opportunity notion \cite{hardt2016equality} requires the same True Positive Rate across different groups. Aristotle regarded justice as “equals should be treated equally and unequals unequally”  (Aristotle 1958). In essence, fairness in machine learning requires that “equal” individuals/groups should be treated “equally”. The key issue is how to establish the similarity criteria to define what are ``equal individuals/groups'' and how to determine what kind of equality treatment is appropriate in certain situations. In light of this, various of fairness notions have been proposed, which we will introduce in more depth in the following sections.


\begin{figure}[t]
\centering\small
\includegraphics[scale=0.6]{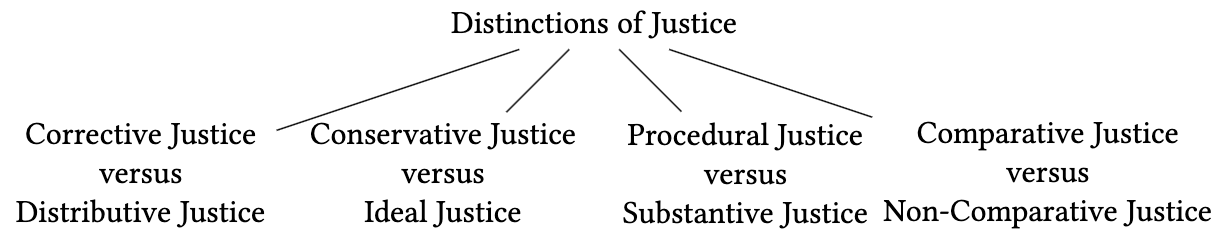}
\caption{Hierarchy of Distinctions of Justice}
\label{justicefig}
\end{figure}

\subsection{The Distinctions of Justice}
In this section, we briefly introduce the four distinctions of justice as analyzed in \cite{miller2021, tang2022and}, and their connections with fairness considerations in machine learning. The hierarchy of the distinctions of justice are shown in Figure.\ref{justicefig}.

The first distinction is \textit{Conservative Justice} versus \textit{Ideal Justice}. \textit{Conservative Justice} interprets justice as solely concerned with what individuals can claim in accordance with existing laws and social conventions, and how individuals can address the injustice in the present world to strive for a better future \cite{miller2021}. \textit{Ideal Justice} views justice on the opposite hand, which demands proposing principles to assess, reform or even abolish the existing norms and practices in order to create a world that is perfectly just \cite{miller2021}. Most of the fairness considerations in machine learning follow the idea of ideal methodology to propose strict fairness requirements and build ideally fair systems \cite{tang2022and}. In specific, current fairness works in machine learning usually first put forward a fairness requirement as the law of a fair world, such as the Equal Opportunity \cite{hardt2016equality}, and then propose methods to meet the fairness requirement, which is typically accompanied with the loss of model performance. However, few works have noticed that the current ideal methodology based methods have essential shortcomings and encourage the study on conservative (non-ideal) perspective for enhancing algorithmic fairness \cite{fazelpour2020algorithmic}. In some areas, building ideally fair systems may even be harmful, for example, \citeauthor{zhang2022improving}~\cite{zhang2022improving} study the fairness of chest X-ray classifiers. The authors emphasize that the group fairness notions often require correcting the performance disparities between advantaged and disadvantaged groups, but not all gaps need to be eliminated in healthcare since some tasks might be inherently more difficult for some groups, e.g., the survival predictions for elderly cancer patients. Consideration of fairness from a conservative and non-ideal perspective is crucial and challenging, which requires the extensive domain-knowledge to understand the reason for unfairness, and the dynamic long-term impact of potential fairness-promoting interventions \cite{fazelpour2020algorithmic}.

The second distinction is \textit{Corrective Justice} versus \textit{Distributive Justice}. \textit{Corrective Justice} regards justice as a remedial principle, which applies when one person wrongly interferes with another, while \textit{Distributive Justice} concerns justice as building principles for allocating various types of distributable goods to individuals \cite{miller2021}. Since recommender system can be considered as a type of resource allocation system \cite{zhang2016economic}, fairness considerations in recommendation mostly focus on the distributive justice to allocate items/opportunities among individuals/groups fairly, for example, the equal recommendation quality between male and female users, or the same admission recommendations for two individuals with the same education backgrounds but different races. When machine learning techniques are used in some certain fields such as criminal justice, the corrective perspective might be considered to better fit the application scenario \cite{miller2021, tang2022and}.

The third distinction is \textit{Procedural Justice} versus \textit{Substantive Justice}. \textit{Procedural Justice} emphasizes the justice in the decision-making process, whereas \textit{Substantive Justice} concentrates on the justice of the final decisions \cite{miller2021}. Most of the association-based fairness notions take the substantive idea and require the equal treatment on performance metrics for different groups or individuals, for example, Equal Opportunity \cite{hardt2016equality} requires the same True Positive Rate across different groups, and Equalized Odds \cite{berk2021fairness} requires that different groups should have the same True Positive Rate and False Positive Rate. Recently, to better assess unfairness issues of the machine learning models, causality-based fairness notions have been proposed to reason about the causal relations between the predicted results and input features \cite{makhlouf2020survey}. Methods to achieve causality-based fairness notions usually involve blocking the causal effects from the sensitive features to the predicted outcomes which aligns with the idea of the procedural justice/fairness \cite{makhlouf2020survey,tang2022and}.

The fourth distinction is \textit{Comparative Justice} versus \textit{Non-Comparative Justice}. When establishing what one deserves, \textit{Comparative Justice} necessitates that one looks at the claims of others, while \textit{Non-Comparative Justice} considers only his or her own pertinent characteristics \cite{miller2021}. For fairness notions proposed in machine learning, both comparative and non-comparative views are presented \cite{tang2022and}. Most of the association-based fairness notions consider comparative perspective. Such notions typically entail identifying similar individuals or groups first, and then ask that they be treated similarly \cite{caton2020fairness, wan2021modeling, mehrabi2021survey, castelnovo2022clarification}. For example, the group fairness notions require that the protected groups should be treated similarly as the advantaged groups such as Statistical Parity \cite{dwork2012fairness,zafar2019fairness} and Equalized Odds \cite{berk2021fairness}; and the individual fairness notions require that similar individuals should be treated similarly such as Fairness Through Awareness \cite{dwork2012fairness}. Non-comparative considerations are also delivered to the fairness notions in machine learning, such as Counterfactual Fairness \cite{kusner2017counterfactual}, which requires that the predicted results for any possible individual in the counterfactual world should be the same as in the real world, where the sensitive features such as the race of the given individual are changed in the counterfactual world. Counterfactual fairness emphasizes the independence between the sensitive features and the predicted outcomes in essence, and thus focuses on determining what is owed to people in accordance with solely their true traits.

\subsection{The Scope of Justice}
The scope of justice includes who and when the rules of justice are applied. \citeauthor{miller2021} \cite{miller2021} analyzes some aspects of these questions in detail. For example,  what kinds of creatures should be covered within the scope of justice principles? Some philosophers believed that the concepts of justice should only apply to humans; while, others supported taking the rights of non-human animals into account \cite{miller2021}. The researches on fairness in machine learning and recommender systems primarily consider human requirements for fairness with a few possible exceptions such as applications of machine learning or AI technologies in biological or medical fields, which may necessitate taking into account non-human animal fairness. Although item-side fairness research is an important topic in recommender systems where items are inherently non-human, researchers are essentially concerned about the human interests, such as producer profits, behind the items \cite{ge2021towards, ge2022explainable}. Another crucial question is whether justice is relational or non-relational \cite{miller2021}. Some researchers suggested that the principles of social justice should only apply among individuals who are engaged together in a certain relationship, while other theories offer alternative descriptions that the principles of justice have universal scope \cite{miller2021}. The algorithmic fairness in machine learning mainly focuses on the relational theory of justice and concerns about the unfairness issues under certain scenarios or within certain domains \cite{tang2022and}, for example, whether a recommender system offers different recommendation quality to users A and B, or whether a college admission process favors accepting applicants of a certain race than others. Other interesting questions about the scope of justice involves discussing the duties of individuals when institutions are established to deliver justice, and justice as recognition to avoid the harms from the failures and mistakes of social recognition to individuals \cite{miller2021}. In order to ensure and advance social justice and fairness, machine learning and AI technologies must act more responsibly given their growing social influence.

\section{Fairness in Classification}\label{classification}
To better understand the concepts introduced above, we briefly talk about fairness in classification as an example since it is a basic and important task in machine learning and has been relatively well-studied in fairness research. Moreover, recommendation problems can sometimes be formulated as a classification task, e.g., when we try to predict whether the user will click an item or not.  \citeauthor{pedreshi2008discrimination} \cite{pedreshi2008discrimination} first explored fair classification to avoid discrimination in classification rule mining. In recent years, a flurry of methods are proposed for promoting fairness in classification. In the following, we first talk about certain fairness concerns and measurements in classification task, and then introduce the methods together with examples for solving the unfairness issue.

\subsection{Fairness Concerns}

Without losing generality, we mainly introduce the fairness research in binary classification task here. In a binary classification task, there is training data $\mathcal{D_T}=\left\{\left(\boldsymbol{x}_{i}, y_{i}\right)\right\}_{i=1}^{N}$ with user feature vectors $\boldsymbol{x} \in \mathbb{R}^{d}$ and the corresponding class labels $y \in\{-1,1\}$. The aim of binary classification problem is to predict the label $\hat{y}_{i}$ through learning a mapping function $f_{\boldsymbol{\theta}}\left(\boldsymbol{x}_{i}\right)$. For example, $\hat{y}_{i}= 1$ if $f_{\boldsymbol{\theta}}\left(\boldsymbol{x}_{i}\right) > 0.5$ and $\hat{y}_{i}= -1$ otherwise. In the literature of fairness in binary classification, each user has an associated sensitive feature $s \in\{0,1\}$. The aim of fair classification is to avoid the unethical interference of sensitive features into the decision-making process. To this end, several fairness notions have been proposed to measure the unfairness of classifiers \cite{caton2020fairness, wan2021modeling}. The two basic frameworks in recent studies on fair classification are group fairness and individual fairness. More formally, we introduce the representative fairness notions as follows:

\subsubsection{\textbf{Group fairness}}

Group fairness requires that the protected groups should be treated similarly as the advantaged groups. Most of the fair classification studies focus on group fairness concerns. The group fairness notions include the \textit{predicted positive rate}-based metrics and the \textit{confusion matrix}-based metrics \cite{caton2020fairness}. \textit{Predicted positive rate}-based metrics require the parity of the predicted positive rates $\operatorname{Pr}(\hat{y}=1)$ across different groups, while \textit{confusion matrix}-based metrics take the True Positive Rate (TPR), True Negative Rate (TNR), False Positive Rate (FPR), and False Negative Rate (FNR) into consideration so that the differences between groups can be captured on a more detailed granularity \cite{caton2020fairness}. One example of \textit{Predicted positive rate}-based metrics is:

\textit{Statistical Parity}: The Statistical Parity, also called \textit{Demographic Parity} or \textit{No Disparate Impact}, requires that each group should have the same likelihood to be classified as positive \cite{dwork2012fairness,zafar2019fairness}:
$$
\operatorname{Pr}\left(\hat{y}=1 \mid s = 0\right)=\operatorname{Pr}\left(\hat{y}=1 \mid s = 1\right)
$$

The shortcoming of this notion is that it ignores the difference between groups. For example, the less popular items may be unpopular for a reason, e.g., due to its unsatisfactory quality. As a result, forcefully requiring all of the unpopular items to have the same click through rate as popular items could be unreasonable.
Besides the metrics that focus on the variants of the predicted positive rate $\operatorname{Pr}(\hat{y}=1)$, most of the group fairness criteria are built on confusion matrix. Examples include:

\textit{Equal Opportunity}: The Equal Opportunity fairness requires that the True Positive Rate (TPR) is the same across different groups \cite{hardt2016equality}:
$$
\operatorname{Pr}\left(\hat{y}=1 \mid y=1 , s = 1\right)=\operatorname{Pr}\left(\hat{y}=1 \mid y=1 , s  = 0\right)
$$

\textit{Equalized Odds}: Stricter than Equal Opportunity, the Equalized Odds fairness also takes False Positive Rate (FPR) into account and requires that different groups should have the same true positive rate and false positive rate \cite{berk2021fairness}:
\begin{equation*}
\begin{split}
    & \operatorname{Pr}\left(\hat{y}=1 \mid y=1 , s = 1\right)=\operatorname{Pr}\left(\hat{y}=1 \mid y=1 , s = 0\right)  \\
    \& & \operatorname{Pr}\left(\hat{y}=1 \mid y=-1 , s = 1\right)=\operatorname{Pr}\left(\hat{y}=1 \mid y=-1 , s = 0\right)
\end{split}
\end{equation*}


\textit{Overall Accuracy Equality}: The Overall Accuracy Equality fairness requires the same accuracy across groups \cite{berk2021fairness}:
\begin{equation*}
    \begin{split}
        &\operatorname{Pr}\left(\hat{y}=-1 \mid y=-1 , s = 1 \right)+\operatorname{Pr}\left(\hat{y}=1 \mid y=1 , s = 1 \right)\\
        =&\operatorname{Pr}\left(\hat{y}=-1 \mid y=-1 , s = 0 \right)+\operatorname{Pr}\left(\hat{y}=1 \mid y=1 , s = 0 \right)
    \end{split}
\end{equation*}

\textit{Equalizing Disincentives}: The Equalizing Disincentives fairness requires the same difference between the True Positive Rate (TPR) and False Positive Rate (FPR) across groups \cite{jung2020fair}:
\begin{equation*}
    \begin{split}
        & \operatorname{Pr}\left(\hat{y}=1 \mid y=1 , s = 1\right)-\operatorname{Pr}\left(\hat{y}=1 \mid y=-1 , s = 1\right) \\
        = & \operatorname{Pr}\left(\hat{y}=1 \mid y=1 , s = 0\right)-\operatorname{Pr}\left(\hat{y}=1 \mid y=-1 , s = 0\right)
    \end{split}
\end{equation*}

\textit{Treatment Equality}: The Treatment Equality fairness requires the same ratio of False Negative Rate (FNR) to False Positive Rate (FPR) across groups \cite{berk2021fairness}:
$$
\frac{\operatorname{Pr}\left(\hat{y}=1 \mid y=-1 , s = 1\right)}{\operatorname{Pr}\left(\hat{y}=-1 \mid y=1 , s = 1\right)}=\frac{\operatorname{Pr}\left(\hat{y}=1 \mid y=-1 , s = 0 \right)}{\operatorname{Pr}\left(\hat{y}=-1 \mid y=1 , s = 0\right)}
$$

\subsubsection{\textbf{Individual fairness}}

Instead of considering fairness across groups, individual fairness notions require that similar individuals should be treated similarly \cite{caton2020fairness, wan2021modeling}. Examples include:

\textit{Counterfactual Fairness}: Counterfactual Fairness is an individual-level causal-based fairness notion \cite{kusner2017counterfactual}. It requires that for any possible individual, the predicted result of the learning system should be the same in the counterfactual world as in the real world. Given a set of latent background variables $U$, the predictor $\hat{Y}$ is counterfactually fair if under any context $\boldsymbol{X}=\boldsymbol{x}$ and $S=s$, the equation below holds for all $y$ and for any value $s^{\prime}$ attainable by $S$:
$$
\operatorname{Pr}\left(\hat{Y}_{S \leftarrow s} (U)=y \mid \boldsymbol{X}=\boldsymbol{x}, S=s\right)=\operatorname{Pr}\left(\hat{Y}_{S \leftarrow s^{\prime}}(U)=y \mid \boldsymbol{X}=\boldsymbol{x}, S=s\right)
$$

It is worth noting that counterfactual fairness is a notion defined based on the idea of causality which emphasizes the independence between the sensitive features and the predicted outcomes. However, the techniques to achieve counterfactual fairness can be diverse and are not limited to typical causal inference methods such as interventions, examples include: 
simply removing the sensitive features and their descendants from the model and prediction function
\cite{kusner2017counterfactual}; variational autoencoders \cite{chiappa2019path}; adversarial learning \cite{grari2020adversarial}; data pre-processing \cite{chen2020counterfactual}; causal regularization \cite{di2020counterfactual}; 
data augmentation \cite{ma2022learning}, etc.

\textit{Fairness Through Awareness}: The Fairness Through Awareness requires that any two individuals with similar non-sensitive features should receive similar predicted results \cite{dwork2012fairness}. Let's consider two individuals $\boldsymbol{x}_{1}$ and $\boldsymbol{x}_{2}$. The distance between the two individuals is defined by $d\left(\boldsymbol{x}_{1}, \boldsymbol{x}_{2}\right)$, and the difference of predicted results between the two individuals is computed through $F\left(\hat{y}_{1}, \hat{y}_{2}\right)$. Here, the formulation of $F(\cdot, \cdot)$ and $d(\cdot, \cdot)$ are usually determined by specific tasks. The Fairness Through Awareness requires that:
$$
F\left(\hat{y}_{1}, \hat{y}_{2}\right) \leq \alpha\cdot d\left(\boldsymbol{x}_{1}, \boldsymbol{x}_{2}\right)
$$
which means that the difference in treated should be upper bounded by the difference between the two individuals.


\subsection{Methods}\label{section5.2}
In this section, we introduce the methods for promoting fairness in classification. As we talked above, the techniques of fair classification can mainly be divided into three categories: pre-processing, in-processing and post-processing \cite{zafar2019fairness}. The methods usually first specify one or more fairness notions to achieve and then propose corresponding methods to control the selected notions. 

\subsubsection{\textbf{Pre-processing Method}}
The pre-processing methods focus on mitigating the bias in data so that any model learned from the data will be fair \cite{zafar2019fairness}.
The pre-processing methods can be applied through changing the labels for certain subjects \cite{luong2011k, kamiran2010classification}, mapping the data to a transformed space \cite{calmon2017optimized}, or perturbing the non-sensitive features \cite{feldman2015certifying}. For example, \citeauthor{calders2009building} \cite{calders2009building} study the problem of unfairness in classification and consider the case where there is unjustified dependency between sensitive features and class labels in the input data. They propose two methods to pre-process the training data with the goal of reducing the dependency between sensitive features and class labels while maintaining the overall positive class probability: massaging and re-weighting. For the massaging method, they change the labels of some subjects $x$ with sensitive features $S = s$ from ``$-$'' to ``$+$'', and meanwhile change the labels of the same number of objects with $S \neq s$ from ``$+$'' to ``$-$''. For the re-weighting method, they re-weight the subjects to reduce the dependency. For example, subjects with $S=s$ and class label ``$+$'' will get higher weights than subjects with $S=s$ and class label ``$-$''. Subjects with $S \neq s$ and class label ``$+$'' will get lower weights than subjects with $S \neq s$ and class label ``$-$''. The authors conduct experiments to show that both of the methods can remove the dependency from training data so that the classifier learned from the data would be fairer. However, the accuracy could be sacrificed after the pre-processing.

\subsubsection{\textbf{In-processing Method}}
Most of the works adopt the in-processing methods to modify the training procedure of the classifier \cite{kamishima2012fairness, goh2016satisfying, woodworth2017learning, quadrianto2017recycling}. For example, \citeauthor{zafar2019fairness} \cite{zafar2019fairness} introduce a flexible constraint-based framework to enable the design of fair margin-based classifiers. To design a fair convex boundary-based classifier, the authors propose to minimize the corresponding loss function under fairness constraints during training:
\begin{equation}
    \left.\begin{array}{rl}
\operatorname{minimize} & L(\boldsymbol{\theta}) \\
\text { subject to } & P(. \mid s=0)=P(. \mid s=1)
\end{array}\right\} \text { Classifier loss function }
\end{equation}

Here, the fairness constraints can be replaced with any fairness requirement that we want. Take statistical parity as an example, it requires the following constraint:
\begin{equation}
    \operatorname{Pr}\left(\hat{y}=1 \mid s = 0\right)=\operatorname{Pr}\left(\hat{y}=1 \mid s = 1\right)
\end{equation}

In particular, the authors consider to control the covariance $\operatorname{Cov}_{SP}(s, d_{\boldsymbol{\theta}}(\boldsymbol{x}))$ between the users' sensitive attribute $s$ and the signed distance $d_{\boldsymbol{\theta}}(\boldsymbol{x})$ from the users' feature vector $\boldsymbol{x}$ to the decision boundary,
since this distance decides the value of the predicted label $\hat{y}$.
\begin{equation}
    \operatorname{Cov}_{SP}\left(s, d_{\boldsymbol{\theta}}(\boldsymbol{x})\right)=\mathbb{E}\left[(s-\bar{s}) d_{\boldsymbol{\theta}}(\boldsymbol{x})\right]-\mathbb{E}[(s-\bar{s})] \bar{d}_{\boldsymbol{\theta}}(\boldsymbol{x}) \approx \frac{1}{N} \sum_{(\boldsymbol{x}, s) \in \mathcal{D_T}}(s-\bar{s}) d_{\boldsymbol{\theta}}(\boldsymbol{x})
\end{equation}

To satisfy statistical parity, we need to make sure that the sensitive feature $s$ is irrelevant to the distance $d_{\boldsymbol{\theta}}(\boldsymbol{x})$, and thus the empirical covariance defined above needs to be approximately zero. Therefore, the we can finally get the objective function as bellow:
\begin{equation}
    \begin{array}{ll}
\text { minimize } & L(\boldsymbol{\theta}) \\
\text { subject to } & \frac{1}{N} \sum_{(\boldsymbol{x}, s) \in \mathcal{D_T}}(s-\bar{s}) d_{\boldsymbol{\theta}}(\boldsymbol{x}) \leq c \\
& \frac{1}{N} \sum_{(\boldsymbol{x}, s) \in \mathcal{D_T}}(s-\bar{s}) d_{\boldsymbol{\theta}}(\boldsymbol{x}) \geq-c
\end{array}
\end{equation}

Here, $c \in \mathbb{R}^{+}$is a given threshold to trade off the accuracy and unfairness due to statistical parity. The above optimization problem is convex and the trade-off between the classifier accuracy and fairness constraint is Pareto-optimal. The conducted experiments which simulate statistical parity in classification outcomes show that the fairness constraints could improve fairness, but would suffer sacrifice on accuracy \cite{zafar2019fairness}.

\subsubsection{\textbf{Post-processing Method}}
Post-processing methods focus on modifying the scores learned from the base classifiers so that the new results are fair. For example, the post-processing methods can mitigate unfairness through learning a different decision threshold for a given scoring function or re-weighting the features \cite{hardt2016equality, corbett2017algorithmic, pleiss2017fairness, dwork2018decoupled, menon2018cost}. In \citeauthor{mehrabi2021attributing} \cite{mehrabi2021attributing}, the authors design a post-processing bias mitigation strategy based on attention. The proposed method is flexible and can be used for any fairness notion. First, an attention-based classifier is trained and we can get the attention weight of each input feature. To find out the effect of each feature on the fairness of outcomes, the authors zero out or reduce the attention weight of each feature and measure the difference in fairness of the outcomes based on the desired fairness notion such as Equalized Odds. If the difference on classification accuracy is small but the difference on fairness is large, it indicates that this feature is mostly responsible for unfairness, and it may need to be dropped to mitigate unfairness through reducing its attention weight. The authors can also control the fairness-accuracy trade-off by decreasing the weights of the features that will hurt fairness or accuracy, and increasing the attention weights of the features that can boost accuracy while keeping the fairness.

The pre-processing, in-processing, and post-processing methods have their own benefits and drawbacks, and the choice of method depends on the particular requirements of the problem at hand. In specific, the pre-processing methods may be relatively easy to implement through data manipulation techniques. They also have the benefit of flexibility since they can be applied to any classification algorithm without modifying the exiting models. However, the pre-processing methods may result in the loss of model performance since they may lose some information through modifying the data. Pre-processing methods are suitable when the dataset is easily accessible, modifications are allowed to be made for fairness demands, and when we need to adapt current models for fairness demands without making great changes.

In-processing methods have the pros that they can incorporate specialized fairness objectives during model training, and also provide a more flexible trade-off between fairness and accuracy. However, in-processing methods suffer from the cons that they may be more complex to implement compared with the pre- or post-processing methods as they require modifications to the model themselves. Moreover, as they are designed for specific aims, they are difficult to be generalized to other scenarios. In-processing techniques can be used when we need to build a new model from scratch or when we have the ability to modify the model to incorporate fairness requirements.

Post-processing methods have the model agnostic flexibility as the pre-processing methods. As fairness interventions are separated from the model training, they allow independent updates of fairness requirements. But they might not generate as good performance as pre-processing or in-processing methods since they are limited to change the model output probabilities. Post-processing methods have the benefit when the model is fixed, for example, when we use third-party models.

As for the evaluation of fairness, the metrics usually highly depend on the specific fairness definition or requirement. For example, if we consider statistical parity, we can adopt $SP =|P(\hat{y}=1 \mid s=0)-P(\hat{y}=1 \mid s=1)|$ to evaluate fairness. The datasets for studying fairness in classification usually contain sensitive features, which have been collected and summarized in several previous surveys such as \cite{quy2021survey, mehrabi2021survey}. Therefore, we would not repeat the introduction here, but we will present several datasets for fair recommendation tasks in Section \ref{data}.

\section{Fairness in Ranking}\label{ranking}
Besides fairness studies in classification task, recent works have also raised the question of fairness in rankings. Recommendation algorithms can usually be considered as a type of ranking problem. Therefore, the fairness studies in ranking is pretty enlightening to improve fairness in recommendations. However, the existing works of fair ranking usually only consider unfairness issue from the item side, i.e., the candidates to be ranked, while the concept of fairness in recommender systems has been extended to multiple stakeholders \cite{burke2017multisided}. In this section, we aim to briefly introduce the research on fair ranking and then focus on fairness in recommendation beginning from the next section.

\subsection{Fairness Concerns}

There are two main types of ranking task: \textit{Score-based Ranking} and \textit{Learning to Rank} \cite{zehlike2021fairness}. The difference between score-based ranking and learning to rank is how to obtain the score for ranking. In score-based ranking, the score can be directly computed from a given function, while in learning to rank, the score is estimated by training a model from the preference-enriched training examples. To discuss the fairness works from a machine learning view, we focus on talking about the fairness studies in learning to rank problem.

In the supervised learning to rank task, a set of candidates is given as $\left\{c_{1}, c_{2}, \ldots c_{n}\right\}$, where each candidate is described by a set of features $X$, which may include sensitive features $S$. The goal of a ranking algorithm is to generate a ranking list $l$ which sorts the candidates by their predicted relevance to the search query. The highest-scoring candidates will appear closer to the top of the list and get the most exposure. We typically return the best-ranked $K$ candidates to construct a top-$K$ ranking list. A majority of the existing works on fair ranking focus on list-wise definitions for fairness, which means that the fairness measures depend on the entire list of results for a given query. As we mentioned above, there is also no universal fairness measure in ranking tasks due to the complex system environment and goals. The fairness notions proposed in classification tasks may be adopted in ranking tasks with some additional task-specific considerations. Generally, the fairness definitions in ranking can also be classified into group and individual fairness. However, in this survey, we talk about the taxonomy of fairness in ranking from a more task-specific perspective: \textit{Probability-based Fairness} and \textit{Exposure/Attention-based Fairness} \cite{patro2022fair, zehlike2021fairness}.

\subsubsection{\textbf{Probability-based Fairness}} The probability-based fairness notions in ranking usually require a minimum/maximum number or proportion of the protected candidates in the top-$K$ ranking list \cite{celis2017ranking, zehlike2017fa, asudeh2019designing, geyik2019fairness}. For example, in paper \cite{celis2017ranking}, the fairness is required as:
\begin{equation}
   L_{k \ell} \leq \sum_{1 \leq j \leq k} \sum_{i \in P_{\ell}} x_{i j} \leq U_{k \ell} 
\end{equation}
 where $x$ is a binary assignment matrix and $x_{ij} = 1$ if item $i$ is assigned to position $j$. The upper and lower bound $U_{k \ell}, L_{k \ell} \in \mathbb{Z}_{\geq 0}$ guarantees that a certain number of items with property $\ell$ 
appear in the top-$K$ positions of the ranking. It is worth mentioning that probability-based fairness definitions are proposed for achieving group fairness goals.

\subsubsection{\textbf{Exposure/Attention-based Fairness}}

Another type of fairness definition in ranking is Exposure or Attention-based Fairness \cite{singh2018fairness, morik2020controlling, diaz2020evaluating, zehlike2020reducing}. In fair ranking task, a frequently studied problem is how to distribute the exposure opportunity to candidates fairly since the total exposure of candidates is a limited resource. The competition of exposure comes from the well-known issue in ranking: the position bias \cite{craswell2008experimental}, which means that the expected exposure of candidates or attention from users will reduce significantly as the ranking position increases, and thus a slight difference in estimated relevance could result in a large difference in item exposures. In such setting, the fairness metrics are usually relevant to the exposure of the candidates belonging to different groups, so that the average exposure of groups can be controlled to be proportional to their average relevance to the search query. Different from probability-based fairness notions, the exposure/attention-based fairness can quantify not only group fairness \cite{singh2018fairness, morik2020controlling}, but also individual fairness \cite{biega2018equity, bower2021individually} according to specific formalization. Group fairness can be measured through the difference of the average exposure between different groups $G_{1}$ and $G_{2}$ \cite{zehlike2021fairness}:
$$
F\left(G_{1}, G_{2}\right)=\Big| \frac{1}{\left|G_{1}\right|} \sum_{c \in G_{1}} \text{ Exposure}(c)-\frac{1}{\left|G_{2}\right|} \sum_{c \in G_{2}} \text{Exposure}(c) \Big|
$$

Similarly, individual unfairness can be measured through the difference of exposure between two candidates $c_1$ and $c_2$ \cite{zehlike2021fairness}:
$$
F(c_1, c_2)=|\text{Exposure}(c_1)-\text{Exposure}(c_2)|
$$

\subsection{Methods}
In this section, we introduce the methods for achieving fairness in ranking. Similarly, we also follow the order of pre-processing, in-processing and post-processing. The majority of works of fair ranking adopt in-processing and post-processing methods. The advantages and disadvantages of the three types in ranking are similar as we discussed in Section \ref{2.2}. In this section, we follow the notations of each paper and explain the notions so that readers can easily reference and understand them.

\subsubsection{\textbf{Pre-processing Method}}
Pre-processing methods seek to mitigate bias in training data, so that the models trained from the unbiased data will be fair \cite{lahoti2019ifair, sonoda2021pre}. For example, in \citeauthor{lahoti2019ifair} \cite{lahoti2019ifair}, authors introduce a method for mapping user records into a low-rank representation that reconciles individual fairness through pre-processing. Specifically, given two samples $x_{i}$ and $x_{j}$, and let's use $x^*_{i}$ and $x^*_{j}$ to denote their representations with only non-sensitive features. The method aims to learn a mapping function $\phi$ so that the individuals who are indistinguishable on their non-sensitive features $x^*$ in data should also be indistinguishable in their transformed representations $\tilde{x}$ under a given distance function $d$:
$$
\left|d\left(\phi\left(x_{i}\right), \phi\left(x_{j}\right)\right)-d\left(x_{i}^{*}, x_{j}^{*}\right)\right| \leq \epsilon
$$

To this end, the problem is formalized as a probabilistic clustering problem. Specifically, we aim for $K$ clusters, each given in the form of a prototype vector $v_{k}$ with $k=1 \cdots K$. Every sample $x_i$ is assigned to a cluster according to a probability distribution $u_{i}$ which reflects the distance of the sample $x_i$ from prototypes. Here the $u_{i k}$ represents the probability of $x_{i}$ belonging to the cluster of prototype $v_{k}$. Therefore, the representation $\tilde{x}_{i}$ is given by: 
$$
\tilde{x}_{i}=\phi\left(x_{i}\right)=\sum_{k=1\cdots K} u_{i k} \cdot v_{k}
$$

To ensure the utility of any downstream application, we can optimize the utility objective through minimizing the data loss induced by $\phi$:
$$
L_{u t i l}(X, \tilde{X})=\sum_{i=1}^{M}\left\|x_{i}-\tilde{x}_{i}\right\|_{2}=\sum_{i=1}^{M} \sum_{j=1}^{N}\left(x_{i j}-\tilde{x}_{i j}\right)^{2}
$$

Furthermore, to reduce individual unfairness in data, we also need to optimize the fairness objective as:
$$
L_{f a i r}(X, \tilde{X})=\sum_{i, j=1 \ldots M}\left(d\left(\tilde{x_{i}}, \tilde{x}_{j}\right)-d\left(x_{i}^{*}, x_{j}^{*}\right)\right)^{2}
$$

The final objective combines the utility loss and fairness loss above as:
$$
L=\lambda \cdot L_{u t i l}(X, \tilde{X})+\mu \cdot L_{f a i r}(X, \tilde{X})
$$
where $\lambda$ and $\mu$ are hyper-parameters, and the objective can be optimized through gradient descent. The transformed representation reconciles individual fairness and can be incorporated into a variety of tasks such as classification and learning-to-rank. Experiments on learning-to-rank show that the proposed methods achieve best individual fairness but worse utility than baseline models \cite{lahoti2019ifair}.

\subsubsection{\textbf{In-processing Method}}

Most of the works on fair ranking adopt in-processing methods, which aim to directly learn a fair ranking model from scratch \cite{singh2019policy, zehlike2020reducing,narasimhan2020pairwise}. For example, \citeauthor{narasimhan2020pairwise} \cite{narasimhan2020pairwise} study the problem of learning pairwise fairness for ranking. Suppose there are queries $S$ drawn \textit{i.i.d.} from an underlying distribution $D$, each candidate to be ranked for answering the query is associated with a vector $x \in X$ and a label $y \in Y$. For example, $Y$ may take the value of 1 or 0 to represent whether the result is clicked by a user. The ranking algorithm needs to learn a scoring function $f: X \rightarrow \mathbb{R}$ for sorting the candidates. Suppose there are a set of  $K$ groups $G_{1}, \cdots, G_{K}$ and each example belongs to exactly one group. The authors define the group-dependent pairwise accuracy $A_{G_{i}>G_{j}}$ as the probability of a clicked candidate from group $G_i$ being ranked above another relevant unclicked candidate from group $G_j$:
\begin{equation}
    A_{G_{i}>G_{j}}:=
P\left(f(x)>f\left(x^{\prime}\right) \mid y>y^{\prime},(x, y) \in G_{i},\left(x^{\prime}, y^{\prime}\right) \in G_{j}\right)
\end{equation}
where $(x, y)$ and $\left(x^{\prime}, y^{\prime}\right)$ are drawn $i . i . d$. from the distribution of examples. The cross-group pairwise equal opportunity is defined as:
\begin{equation}
    A_{G_{i}>G_{j}}=\kappa, \text { for some } \kappa \in[0,1] \text { for all } i \neq j.
\end{equation}

The training problem can be formulated as maximizing the overall pairwise accuracy under the cross-group equal opportunity fairness constraint:
\begin{equation}
   \begin{gathered}
\max_{f \in \mathcal{F}} \text{AUC}(f) \\
\text { s.t. } \quad A_{G_{i}>G_{j}}(f)-A_{G_{k}>G_{l}}(f) \leq \epsilon \quad \forall i \neq j, k \neq l .
\end{gathered}
\end{equation}
where $\text{AUC}(f)$ is the overall pairwise accuracy and $\mathcal{F}$ is the class of models we are interested in. Experiment results show that the proposed method can greatly reduce the fairness violation at the cost of a lower test AUC. The paper claims that their algorithm can be applied to any pairwise metric.

\subsubsection{\textbf{Post-processing Method}}
Post-processing methods usually re-rank candidates in the output list of the base ranking models \cite{liu2018personalizing, biega2018equity, zehlike2017fa}. For example, \citeauthor{singh2018fairness} \cite{singh2018fairness} study the problem of fair exposure in rankings. They consider the setting where the relevance of documents has been obtained or well estimated from some base rankers, and they only ask how to fairly allocate exposure in rankings based on the known relevance. For a given single query $q$, the ranking algorithm aims to present a ranking $r$ of a set of documents $\mathcal{D}=\left\{d_{1}, d_{2}, d_{3} \cdots d_{N}\right\}$. The utility of a ranking $r$ for query $q$ is denoted as $\mathrm{U}(r|q)$. The authors propose to optimize the ranking task under fairness constraints so that the learned ranking $r$ maximizes the utility function and fairness. To learn the distribution of the ranking $r$, the authors learn a probabilistic ranking matrix $P$ where $\mathbf{P}_{i, j}$ is the probability that $r$ places document $d_{i}$ at rank $j$. Here, we need both the sum of probabilities for each position and the sum of probabilities for each document to be 1 , i.e. $\sum_{i} \mathbf{P}_{i, j}=1$ and $\sum_{j} \mathbf{P}_{i, j}=1$. Therefore, the problem of optimal fair ranking can be formulated as finding the utility-maximizing probabilistic ranking $P$ under fairness constraints: 
\begin{equation}
   \begin{array}{rlr}
\mathbf{P}=\operatorname{argmax}_{\mathbf{P}} & \mathrm{U}(r|q) & \text { (expected utility) } \\
\text { s.t. } & \mathbbm{1}^{T} \mathbf{P}=\mathbbm{1}^{T} \quad & \text { (sum of probabilities for each position) } \\
& \mathbf{P} \mathbbm{1} = \mathbbm{1} \quad &\text { (sum of probabilities for each document) } \\
& 0 \leq \mathbf{P}_{i, j} \leq 1 \quad &\text { (valid probability) } \\
& \mathbf{P} \text { is fair } & \text { (fairness constraints) }
\end{array}
\end{equation}

The proposed method is able to achieve various fairness notions in ranking depending on specific fairness requirements. Here we also take Statistical Parity of exposure as an example. First of all, the exposure for a document $d_i$ under a probabilistic ranking $P$ is defined as:
\begin{equation}
    \operatorname{Exposure}\left(d_{i}|\mathbf{P}\right)=\sum_{j=1}^{N} \mathbf{P}_{i, j} \mathbf{v}_{j}
\end{equation}
where $v_j$ represents the importance of position $j$ and the exposure for a group $G_k$ is the average exposure of each document in $G_k$:
\begin{equation}
    \operatorname{Exposure}\left(G_{k} | \mathbf{P}\right)=\frac{1}{\left|G_{k}\right|} \sum_{d_{i} \in G_{k}} \operatorname{Exposure}\left(d_{i} | \mathbf{P}\right)
\end{equation}

Based on this, the Statistical Parity constraint requires that:
\begin{equation}
\text{Exposure}\left(G_{0} | \mathbf{P}\right)=\operatorname{Exposure}\left(G_{1} | \mathbf{P}\right)
\end{equation}

After derivation and simplification in this paper, the fairness constraint can be plugged into the linear program to solve. The experiments show that the effect of statistical parity in ranking is similar to its effect in classification, where it can also lead to a drop of accuracy.

The pre-, in-, and post-processing methods in ranking also have their advantages and disadvantages, similar to the pros, cons, and suitable application scenarios we introduced in Section \ref{section5.2}. As a result, the choice of methods needs careful consideration of the specific fairness demands, task objectives, resource availability, and legal requirements.
The evaluation of fairness in ranking is also highly relevant to the specific fairness definition or requirement. The survey \cite{zehlike2021fairness} summarized some datasets for studying fairness in ranking.


\section{Fairness in Recommendation}\label{rec}

In this section, we introduce the definitions, methods, evaluation, datasets and challenges for fairness in recommendation research. We first briefly talk about the preliminaries of recommender systems and show several examples of unfairness in recommendation. After that, we systematically show the taxonomy of fairness notions, techniques of promoting fairness, evaluating fairness, and datasets for studying fairness in recommendation. We also analyze the challenges and opportunities in fair recommendation field to inspire more works in the future. 

\subsection{Preliminaries of Recommender Systems}

In recommendation task, there is usually a user set $\mathbb{U}=\left\{u_{1}, u_{2}, \cdots, u_{n}\right\}$ and an item set $\mathbb{V}=\left\{v_{1}, v_{2}, \cdots, v_{m}\right\}$, where $n$ is the number of users and $m$ is the number of items. We also have the user-item interaction histories represented as a 0-1 binary matrix $H =\left[h_{i j}\right]_{n \times m}$, where each entry $h_{ij}=1$ if user $u_i$ has interacted with item $v_j$, otherwise $h_{ij}=0$. The main task for recommendation is to predict the preference scores of users over items $S_{uv}$, so that the model can recommend each user $u_{i}$ a top-$N$ recommendation list $\left\{v_{1}, v_{2}, \cdots, v_{N} | u_{i}\right\}$ according to the predicted scores. To learn preference scores, modern recommender models usually learn user and item representations from data, and then take the representations as input to a learned or designed scoring functions to make recommendations. Most of the Collaborative Filtering (CF) \cite{goldberg1992using,sarwar2001item,ekstrand2011collaborative} or Collaborative Reasoning (CR) \cite{chen2021neural,shi2020neural,chen2022graph} recommender systems are directly trained from user-item interaction history. The content-based recommendation models \cite{adomavicius2011context,lops2011content,zhang2017joint} and hybrid models \cite{burke2002hybrid} may also leverage user/item profiles as input or use additional information to help train the model. In fair recommendation works, fairness concerns are usually relevant to user and item features such as user gender or race, and item category or popularity. 

\begin{figure}
        \centering
        \pgfplotsset{label style={font=\Large},
                    tick label style={font=\Large}}

        \begin{subfigure}[b]{0.35\textwidth}
            \centering
            \resizebox{1\textwidth}{!}{
            \includegraphics[scale=0.5]{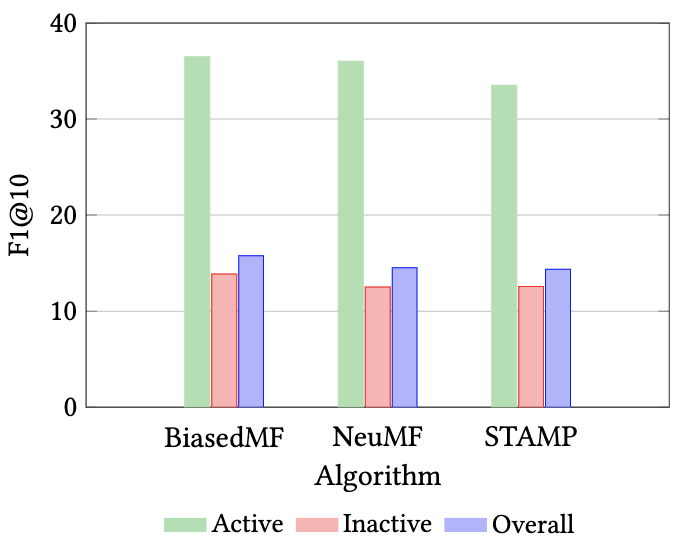}
            }
            \caption[]%
            {{\small User-side unfairness}}    
            \label{fig:user-fair}
        \end{subfigure}
        \begin{subfigure}[b]{0.35\textwidth}
            \centering 
            \resizebox{1\textwidth}{!}{
                \includegraphics[scale=0.5]{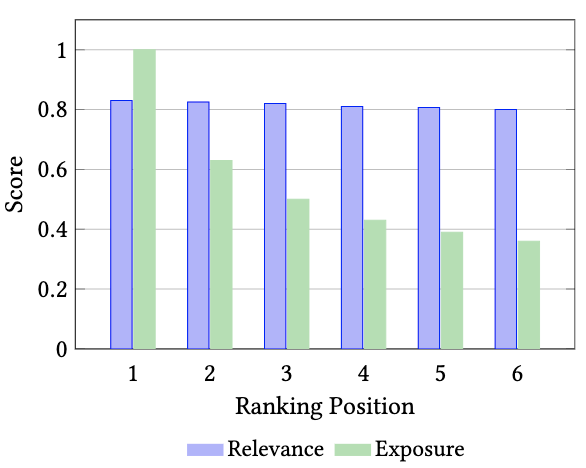}
            }
            \caption[]%
            {{\small Item-side unfairness}}    
            \label{fig:item-fair}
        \end{subfigure}
        \caption[]{\small (a) The significant difference between inactive and active user groups on recommendation quality, image from \cite{li2021user}; (b) The significant exposure difference among items with close relevance scores.} 
        \label{compare}
        \vspace{-10pt}
\label{fig1}
\end{figure}

\subsection{Examples of Unfairness in Recommendation}

The fairness considerations can be raised from very different perspectives in recommendation task. We show two examples of unfairness in recommendation in Fig. \ref{fig1}, one from user-side and another from item-side. Fig.(\ref{fig:user-fair}) shows the unfairness of recommendation quality between groups of active users and inactive users in an e-commerce dataset \cite{li2021user}. We first sort users according to their number of interactions in the training data. We then label the top 5\% users as active ones and leave the remaining 95\% users into the inactive group. After that, we test the performance of three fairness-unaware recommendation models (BiasedMF \cite{koren2009matrix}, a shallow model, NeuMF \cite{he2017neural}, a deep model, and STAMP \cite{liu2018stamp}, a sequential model) on two user groups. We can see that the recommendation quality (F1 score) of active users is significantly higher than those inactive ones, which shows that the recommendation models are dominated by the active users and thus generate unfairness treatments for inactive users, which also degrade the overall performance since the inactive users are the majority. Similar phenomenon is also observed on some other datasets \cite{boratto2022consumer,rahmani2022experiments}.
In addition, Fig.(\ref{fig:item-fair}) shows the unfairness of item exposure in recommendation lists. Suppose there are a set of items being recommended to a user and each item has a learned relevance score. The recommender systems usually sort the items by their relevance scores and the highly scored items will appear closer to the top of the recommendation list. Here we consider a standard exposure drop-off which is commonly used in the Discounted Cumulative Gain measure: $1 / \log (1+i)$ to compute the exposure of each position, where $i$ is the position in the list \cite{singh2018fairness}. We can see from the figure that although the items have very close relevance scores, the exposure opportunity received by each item significantly varies, which causes unfairness of exposure for items in recommendation. A similar example is also discussed in \cite{singh2018fairness}.





\begin{table}[t]
    \centering
    \small
    \setlength{\tabcolsep}{2.5pt}
    \caption{The Taxonomy of Fairness Notions in Recommender Systems.}
    \begin{tabular}{ccc}
    \toprule
    \textbf{Criterion} & \textbf{Category} & \textbf{Reference}\\
    \midrule
        \multirow{2}[2]{*}{Level} & Group &
        \cite{zhu2018fairness} \cite{mehrotra2018towards} \cite{zhu2018fairnessworkshop} \cite{geyik2019fairness} \cite{nandy2020achieving} \cite{li2021user} \cite{ge2021towards} \cite{islam2021debiasing} \cite{tsioutsiouliklis2022link} \cite{naghiaei2022cpfair} \cite{ge2022toward} \cite{chen2023improving} \cite{chen2023fairly} \\
                                & Individual & \cite{patro2020fairrec}
                     \cite{marras2021equality} \cite{li2021counterfactual} \cite{li2022fairgan} \cite{do2021online} \cite{li2023transferable} \\
    \midrule
        \multirow{2}[2]{*}{Subject*} & User &  \cite{fu2020fairness} \cite{kaya2020ensuring} \cite{li2021counterfactual} \cite{islam2021debiasing} \cite{wu2021learning} \cite{wu2022selective} \cite{sato2022enumerating} \cite{do2021online}  \cite{wu2022fairrank} \cite{wu2022big} \cite{li2023transferable} \cite{lifairness}\\
        & Item &
         \cite{beutel2019fairness}
        \cite{nandy2020achieving}
        \cite{zhu2021fairness} \cite{gomez2021winner}
        \cite{shen2021sar}
        \cite{wang2021practical}
        \cite{li2022fairgan} \cite{qi2022profairrec} \cite{yang2022effective}\cite{xu2023p} \cite{wu2023faster} \cite{zhang2023asymmetrical}\\
    \midrule
        \multirow{2}[2]{*}{Side*} & Single-side & \cite{leonhardt2018user} \cite{fu2020fairness} \cite{zhu2021fairness} \cite{gomez2021winner} \cite{li2021user} \cite{li2021counterfactual} \cite{li2022fairgan} \cite{islam2021debiasing} \cite{wu2021learning}  \cite{wu2021fairrec} \cite{wu2022big}  \cite{qi2022profairrec} \\
        & Multi-side &
        \cite{burke2018balanced}
        \cite{abdollahpouri2019multi} \cite{patro2020fairrec}
        \cite{patro2020incremental}
        \cite{wu2021tfrom} \cite{wu2021multi} \cite{biswas2021toward} \cite{mondal2020two}
        \cite{wu2022joint} \cite{naghiaei2022cpfair} \cite{liu2023mitigating} \cite{smith2023many}\\
    \midrule
        \multirow{2}[2]{*}{Relationship} & Associative & \cite{yao2017beyond} \cite{farnadi2018fairness} \cite{zhu2018fairness} \cite{chakraborty2019equality} \cite{bobadilla2020deepfair} \cite{li2022fairgan} \cite{boratto2021interplay} \cite{zhu2021fairness} \cite{wu2022big} \cite{rastegarpanah2019fighting} \cite{chaudhari2020general} \cite{li2021user} \\
        & Causal & \cite{li2021counterfactual} \cite{huang2021achieving} \cite{wang2021deconfounded}\\
    \midrule
        \multirow{2}[2]{*}{State*} & Static &  \cite{yao2017beyond}  \cite{xiao2017fairness} \cite{resheff2018privacy} \cite{leonhardt2018user} \cite{beutel2019fairness} \cite{fu2020fairness} \cite{chaudhari2020general} \cite{boratto2021interplay} \cite{li2021user} \cite{li2021counterfactual} \cite{polyzou2021faireo} \cite{dong2021user} \\
        & Dynamic & \cite{ge2021towards} \cite{liu2021balancing} \cite{akpinar2022long} \\
    \midrule
        \multirow{2}[2]{*}{Duration*} & Short-term & 
        \cite{yao2017beyond} \cite{xiao2017fairness} \cite{resheff2018privacy} \cite{leonhardt2018user} \cite{kamishima2018recommendation}
        \cite{beutel2019fairness}
        \cite{abdollahpouri2019unfairness}
        \cite{fu2020fairness}
        \cite{abdollahpouri2020connection}
        \cite{li2021user} \cite{li2021counterfactual} \cite{tsioutsiouliklis2022link} \\
        & Long-term &
        \cite{borges2019enhancing} \cite{ge2021towards}
        \cite{biswas2021toward} \cite{liu2021balancing}
        \cite{fabbri2021exposure} \cite{akpinar2022long} \cite{ge2022toward} \\
    \midrule
        \multirow{2}[2]{*}{Granularity*} & Populational  &  \cite{borges2019enhancing}
         \cite{fu2020fairness}
        \cite{li2021user} \cite{ge2021towards} \cite{zhu2021fairness1}
        \cite{liu2021balancing}
        \cite{gomez2021winner} \cite{wu2022big}  \cite{qi2022profairrec}
        \cite{wu2022joint} \cite{naghiaei2022cpfair}
        \cite{akpinar2022long}
       
         \\
        & Personalized & \cite{li2021counterfactual} \cite{wu2022selective} \\
    \midrule
        \multirow{2}[2]{*}{Transparency} & Blackbox & 
        \cite{yao2017beyond} \cite{farnadi2018fairness}
        \cite{li2021counterfactual} \cite{huang2021achieving}
         \cite{fabbri2021exposure}
         \cite{li2021user}
         \cite{zhu2021fairness}
         \cite{wu2021tfrom} \cite{wu2021multi} \cite{biswas2021toward}
         \cite{akpinar2022long} \cite{ge2022toward} 
        
         \\
        & Explainable & \cite{ge2022explainable} \\
    \midrule
        \multirow{2}[2]{*}{Centrality*} & Centralized & \cite{leonhardt2018user} \cite{kaya2020ensuring} \cite{li2021user} \cite{li2021counterfactual} \cite{islam2021debiasing} \cite{wu2021learning}  \cite{zhu2018fairness} \cite{mehrotra2018towards} \cite{beutel2019fairness} \cite{geyik2019fairness}
        \cite{nandy2020achieving} \cite{gomez2021winner}   \\
        & Federated & \cite{liu2022fairness} \cite{zhu2022cali3f}\\
    \bottomrule
    \end{tabular}
    \label{tab:taxonomy}
\end{table}

\subsection{A Taxonomy of Fairness Notions in Recommendation}
In this section, we introduce the taxonomy of fairness notions in recommendation to provide a systematic overview of the various perspectives of fairness in recommendation \cite{li2021tutorial, li2021cikm}. In Table.\ref{tab:taxonomy}, we summarize the categories as well as several corresponding papers in each category. It is worth noting that some fairness notions such as the \textit{User Fairness}, \textit{Multi-sided Fairness}, and \textit{Dynamic Fairness} might be special considerations in recommendation scenarios, while some fairness notions such as \textit{Group Fairness}, \textit{Individual Fairness} and \textit{Causal Fairness} can be considered similarly in various machine learning tasks such as classification. To make a differentiation, the criterion with star in Table.\ref{tab:taxonomy} are those proposed considering the special characteristics of recommendation tasks such as the interactivity, multi-sidedness, and dynamics, while other criterion can be considered more generally in diverse machine learning tasks. In this section, in addition to define each category formally,  we also introduce representative works of each category in recommendation.

\subsubsection{\textbf{Group Fairness vs. Individual Fairness}}
As we introduced above, group fairness and individual fairness are two basic frameworks in algorithmic fairness and the fairness in recommendation can also be divided into these two categories.

\begin{itemize}
   \item
    \textbf{Group Fairness}: Group fairness requires that different predefined groups should be treated equally.
   \item
   \textbf{Individual Fairness}: Individual fairness believes that similar individuals should receive similar treatments.
\end{itemize}

Here we introduce some representative examples for each category. \citeauthor{yao2017beyond} \cite{yao2017beyond} present four group fairness metrics in collaborative filtering recommender systems. The authors consider a binary group feature such as gender to divide disadvantaged and advantaged groups and all proposed metrics measure a discrepancy between the prediction behavior for disadvantaged users and advantaged users. \citeauthor{li2021user} \cite{li2021user} consider user-oriented group fairness in commercial recommendation. The authors divide users into advantaged and disadvantaged groups according to their activity, e.g., the number of user interactions in the training data, and
require that a fair recommendation algorithm should offer similar recommendation quality for different groups of users. The authors use the difference of the average recommendation performance (such as F1 and NDCG) between two groups to measure the user group unfairness of a recommendation algorithm. 
Results show that fairness-aware learning can achieve both better fairness on recommendation quality between the two groups and better overall recommendation performance of the system, which means that when properly treated, fairness and utility may not conflict with each other and they can sometimes be improved simultaneously. 
Similar results are also observes on more datasets and recommendation scenarios \cite{rahmani2022experiments}. Furthermore, \citeauthor{rahmani2022experiments} \cite{rahmani2022experiments} show that the recommendation accuracy disparity among user groups may vary on different datasets. In general, such disparity is more significant in recommendation scenarios where user behaviors require significant marginal cost, such as e-commerce where users really have to pay money to purchase a product and POI recommendation where users have to spend travel costs (both time and money) to check in a location. In these scenarios, the behavioral difference between advantaged and disadvantaged users can be more apparent due to their different ability to afford the costs. However, such disparity can be less severe in some other recommendation scenarios such as music recommendation, since once users have the membership, the marginal cost of listening a song is relatively small (just a few minutes of time). This shows that it is important to carefully define the user groups according to the recommendation scenario when considering group fairness in recommendation.

Some other works consider individual fairness in recommendation. For example, \citeauthor{xiao2017fairness} \cite{xiao2017fairness} study the problem of group recommendation where items are recommended to groups of users whose preferences can be different from each other. The authors define the individual utility given a recommendation to the group by considering the relevance of each recommended item to the user, and treat the imbalances between group members' utilities as the metric of fairness. Authors maximize the satisfaction of each group member while minimize the unfairness between them. \citeauthor{li2021counterfactual} \cite{li2021counterfactual} consider counterfactual fairness in recommendation, which requires that the recommendation results for each user are the same in the factual and the counterfactual world. The counterfactual world is defined as the one where user's sensitive features were changed while all the other features that are not causally-dependent on the sensitive features remain the same.

\subsubsection{\textbf{User Fairness vs. Item Fairness}}

As a multi-stakeholder system, fairness requirements in recommender systems may come from different sides including but not limited to user-side and item-side, where the user-side refers to the users who receive recommendations and item-side refers to the items to be ranked or recommended. In some cases such as job applicant recommendation for recruiters, the job seekers are those to be ranked and recommended, but we still consider the fairness demands of them as item-side fairness for simplicity, since even though the job applicants are human users of the job matching system, they are the ``items'' to be ranked from the recommender system's point of view. 

\begin{itemize}
   \item
    \textbf{User Fairness}: A fair recommender system for users should treat different predefined groups of users or similar individual users fairly. 
   \item
   \textbf{Item Fairness}: A fair recommender system for items should treat different predefined groups of items or similar individual items fairly. 
\end{itemize}

The fairness demands from user side are usually about the quality of the recommendations for them, while the fairness considerations from item side usually focus on the exposure opportunity of items in the ranking lists. Here we discuss some examples in the following. 
\citeauthor{fu2020fairness} \cite{fu2020fairness} considers mitigating the unfairness problem for users in the context of explainable recommendation over knowledge graphs. The authors find that performance disparity exists between active and inactive user groups and claim that such disparity may come from the different distribution of path diversity. \citeauthor{wu2022big} \cite{wu2022big} investigate the problem that big recommendation models are unfair to cold-start users. Specifically, the authors observe that the recommendation accuracy of cold-start users is sacrificed during the process of optimizing the overall performance of big recommendation models. The authors proposed a self-distillation approach to encourage the model to fairly capture the interest distributions of both cold-start and active users. Compared with the works considering user-side fairness, some other research focus on item-side fairness. One typical problem of item fairness in recommendation is to mitigate the exposure unfairness which is usually due to popularity bias, i.e., the popular items may get disproportionately more exposure opportunity since they dominate the embedding learning in recommender systems, while the less popular items may get much less exposure opportunity even though they have equally good or even better quality.
The exposure unfairness issue is often solved by increasing the number of unpopular items (long-tail items) or otherwise the overall catalog coverage in the recommendation list, or making sure that the exposure rate of items is proportional to their quality \cite{adomavicius2011improving,kamishima2014correcting,abdollahpouri2017controlling,abdollahpouri2019managing}. Another example of item-side fairness in recommendation is \cite{beutel2019fairness} which proposes the notion of pairwise fairness in recommendation. The authors measure item fairness based on pairwise comparisons, and require that the likelihood of a clicked item being ranked above another relevant unclicked item should be the same across different item groups, conditioned on that the items have been engaged with the same amount.

\subsubsection{\textbf{Single-side Fairness vs. Multi-side Fairness}}
Since the concept of fairness in recommender systems has been extended to multiple stakeholders \cite{burke2017multisided}, it is not satisfactory to meet the fairness demands of only user side or item side, and thus an important problem is how to balance the fairness demands of multiple sides.

\begin{itemize}
   \item
    \textbf{Single-side Fairness}: A single-side fair recommender system only considers fairness demands from one single side, being either user-side, item-side, or platform-side.
   \item
   \textbf{Multi-side Fairness}: A multi-side fair recommender system considers fairness demands from multiple sides at the same time.
\end{itemize}

Therefore, besides the single-side fairness considerations as we mentioned above, several works study multi-side fairness in recommender systems. Both \cite{burke2017multisided} and \cite{abdollahpouri2019multi} present several most commonly observed classes of multi-stakeholder recommendation, and categorize different types of fairness that are important to address in a multi-stakeholder recommendation environment. \citeauthor{patro2020fairrec} \cite{patro2020fairrec} address individual fairness for both producers and customers. From the item-side, authors require to reduce the exposure inequality among items, and from the user-side, authors argue that the platforms should fairly distribute the loss in utility among all the customers. Specifically, the authors formulate the problem as an allocation problem that guarantees minimum exposure for the items and envy-free up to one item (EF-1) \cite{budish2011combinatorial} for the users. Here the EF-1 ensures that every user values his or her allocation at least as much as any other agent's allocation after hypothetically removing the most valuable item from the other agent's allocated bundle. \citeauthor{patro2020incremental} \cite{patro2020incremental} study the scenario where the online recommendation algorithms are updated frequently to improve user utility. The authors argue for incremental updates of the platform algorithms to avoid the abrupt change of item exposures. To ensure the fairness for two-sided platforms, the authors propose an online optimization method to achieve smooth transition of the item exposures while guaranteeing a minimum utility for every customer. From the item side, the authors require a minimal difference between the exposure distribution in the new system and that in the old system, while for the user side, the authors define a platform to be fair for the customers if it guarantees a minimum utility for everyone. \citeauthor{liu2023mitigating}\cite{liu2023mitigating} propose a fairness-centric model and design explicit and implicit discriminators to
mitigate the popularity bias from both the user and item side locally and globally. The extensive experiments show that the proposed method achieves the best fairness performance while maintaining the competitive model effectiveness in recommendation.

\subsubsection{\textbf{Associative Fairness vs. Causal Fairness}}\label{causal}

The fairness notions in machine learning and recommendation tasks can be proposed based on associative or causal perspectives.

\begin{itemize}
   \item
    \textbf{Associative Fairness}: Associative fairness metrics are developed based on measuring the associations or correlations between the predictive outcomes and the sensitive features. 
   \item
   \textbf{Causal Fairness}: Causal fairness metrics are developed based on measuring the causal effects of the sensitive features on the predictive outcomes. 
\end{itemize}

The research community first studied fairness in machine learning by developing association-based (or correlation-based) fairness notions, with the aim to find the discrepancy of statistical metrics between individuals or sub-populations. Up to now, most of the existing works about fairness in recommendations consider the association-based fairness notions. For example, the previously mentioned research \cite{yao2017beyond}, which proposes metrics to measure the discrepancy between the prediction behavior for disadvantaged users and advantaged users in collaborative filtering recommender systems. Recently, researchers have found that fairness cannot be well assessed only based on association notion \cite{khademi2019fairness,kusner2017counterfactual,zhang2018equality,zhang2018fairness}, since they cannot reason about the causal relations among features. However, unfair treatments usually result from a causal relation between the sensitive features (e.g. gender) and model decisions (e.g. admission). Therefore, researchers have proposed causal-based fairness notions~\cite{wu2019pc,kusner2017counterfactual} to study unfairness issues in machine learning more properly. Unlike the association-based notions which are only computed based on data, the causal-based fairness notions also consider the additional structural knowledge of the system regarding how variables propagate on a causal model (e.g., a causal graph) \cite{makhlouf2020survey}. The causal-based fairness notions are usually defined in terms of interventions and counterfactuals \cite{pearl2009causality}. The non-observable properties of interventions and counterfactuals bring great challenges to the applicability of causal-based notions in real world since sometimes they cannot be easily computed from observational data \cite{makhlouf2020survey}. Two main frameworks are proposed to solve the problem: one is known as the potential outcome framework \cite{imbens2015causal}, which usually estimates the causal quantities through re-weighting and matching techniques; another is known as the structural causal model \cite{pearl2009causality}, which tells how to estimate causal quantities through observable data under identifiability criterion \cite{shpitser2012identification}. \citeauthor{li2021counterfactual} \cite{li2021counterfactual} aim to achieve counterfactual fairness in recommender systems. The authors define counterfactually fair recommendation as the recommendation results that are the same in factual and counterfactual world for each possible user. In the paper, the counterfactual world could be the one where user's sensitive features are changed, for example, the gender of a user is changed from male to female, while all other insensitive features which are not causally-dependent on sensitive features remain the same. Compared with the classification problem, causality in fair recommendation has been rarely studied. Since it is now widely accepted that causal-based notions are important to be considered to enhance fairness \cite{makhlouf2020survey}, we believe causality considerations will open up new challenges and opportunities for studying fairness in recommendation.

\subsubsection{\textbf{Static Fairness vs. Dynamic Fairness}}
In recommendation tasks, fairness can be considered from a static perspective or dynamic perspective.

\begin{itemize}
   \item
    \textbf{Static Fairness}: A static fair recommender system considers achieving fairness in a static environment where user preferences and item properties are assumed to be unchanging during the whole recommendation process.
   \item
   \textbf{Dynamic Fairness}: A dynamic fair recommender system considers achieving fairness in a dynamic environment where the user preferences and item properties may change with time during the recommendation process.
\end{itemize}

Most of the existing works about fairness in recommendation consider static fairness where the recommendation environment is fixed during the recommendation process, and usually provide a one-time fairness solution based on fairness-constrained optimization. For example, \citeauthor{li2021user} \cite{li2021user} divide users into advantaged and disadvantaged groups based on their activity and propose a re-ranking method to mitigate the performance disparity between the two groups. The user activity level is assumed to be unchanged and fixed during the recommendation process. However, recommender systems are usually dynamic systems since users constantly interact with items, as a result, a previously inactive user may now become active while a previously unpopular item may now become popular, and vice versa
\cite{mansoury2020feedback}. What's more, researches have shown that imposing static fairness criteria myopically at every step may actually exacerbate unfairness \cite{williams2019dynamic, creager2020causal, d2020fairness, zhang2020fair}. To solve the problem, a few works have paid attention to the dynamic factors in the recommender system environment and study how to enhance fairness with dynamics such as the change of utility, attributes and group labels due to the user interactions throughout the recommendation process \cite{zhang2021recommendation}. \citeauthor{ge2021towards} \cite{ge2021towards} study the dynamic fairness of item exposure in recommender systems. The items are separated into advantaged and disadvantaged groups based on item popularity. However, the item popularity may change during the recommendation process based on the recommendation strategy and user feedback, causing the underlying group labels to change over time. To solve the challenge, the authors formulate the problem as a Constrained Markov Decision Process (CMDP) by dynamically constraining the fairness of item exposure at each iteration. \citeauthor{liu2021balancing} \cite{liu2021balancing} study the task of Interactive Recommender Systems (IRS) where items are recommended to users consecutively and the user feedback is received during the process. IRS gradually refine the recommendation policy according to the obtained user feedback in an online manner and aim to maximize the total utility over the whole interaction period. During the recommendation process, the user preferences and the system's fairness status constantly change over time. To resolve this problem, the authors propose a reinforcement learning based framework, which jointly represents the user preferences and the system's fairness status into the states of the Markov Decision Process (MDP) for recommendation, so that they can dynamically achieve a long-term balance between accuracy and fairness in IRS.

\subsubsection{\textbf{Short-term Fairness vs. Long-term Fairness}}
The fairness of recommender systems can also be considered from short-term or long-term fairness requirements.

\begin{itemize}
   \item
    \textbf{Short-term Fairness}: A short-term fair recommender system only considers achieving fairness at present. 
   \item
   \textbf{Long-term Fairness}: A long-term fair recommender system considers achieving fairness in the long run.
\end{itemize}

The short-term fairness is usually a static fairness since it only needs to consider the current status and achieve fairness for one-time. The dynamic fairness can be seen as a type of long-term fairness since the dynamics usually occur in the long run. However, the long-term fairness covers a broader scope such as the case where there is no dynamics but the fairness cannot be realized in the present and can only be addressed by a long term strategy. For example,  \citeauthor{borges2019enhancing} \cite{borges2019enhancing} consider to achieve the individual exposure fairness for items. In recommendation task, items are presented as an ordered list. For items with very close or equal relevance scores, the algorithm needs to arrange them in a proper order. However, users usually pay more attention to the top positions of the list and the level of attention significantly decreases as the position in the ranking gets lower. In such case, individual exposure fairness (i.e., items with close relevance scores get similar amounts of exposure) cannot be achieved within just one search result due to the limited positions and can only be achieved in the long term by changing the position of items in multiple rounds of ranking so as to achieve fairness in expectation. Specifically, the authors require that the ranked items receive exposure that is proportional to their relevance scores in a series of rankings through an amortized manner.

\subsubsection{\textbf{Populational Fairness vs. Personalized Fairness}}
Most fairness definitions consider fairness on populational level which aim to achieve the same fairness definition for all users. However, users' fairness demands can be very personalized. 
For example, some users may be very sensitive on gender and they want to be fairly treated on the gender feature, while some other users may not care about gender too much but instead they are very sensitive to age and they do not want to be discriminated by age.

\begin{itemize}
   \item
    \textbf{Populational Fairness}: Populational fairness considers the same fairness demand for all the individuals or predefined groups in the whole population. 
   \item
   \textbf{Personalized Fairness}: Personalized fairness considers different and personalized fairness demands of individuals or predefined groups in the population. 
\end{itemize}

It is important to understand that populational vs. personalized fairness is different from group vs. individual fairness. The group vs. individual fairness dimension emphasizes whether users are treated as a group or individually, while populational vs. personalized fairness dimension emphasizes whether users have the right to tell the system what fairness consideration that they care about rather than the system designer care about. More specifically, a system guarantees individual fairness does not necessarily mean that it provides personalized fairness---it could be populational. For example, the system may globally apply the same individual fairness such as counterfactual gender fairness to all users, even though some users do not care about gender fairness but care about age fairness instead. 
Similarity, a system guarantees group fairness does not necessarily mean that it is populational fairness---it could be personalized. For example, the system may split users into two groups and guarantee that the average user satisfaction of group one is fair compared to that of group two, meanwhile, the fairness of each user is defined based on the user's own provided feature. \citeauthor{bose2019compositional} \cite{bose2019compositional} introduced compositional fairness for multiple sensitive attribute combinations for the knowledge graph embedding task. \citeauthor{li2021counterfactual} \cite{li2021counterfactual} further defined personalized fairness based on causal notion for the collaborative filtering recommendation task. The system guarantees personalized counterfactual fairness for each user, i.e., the user's recommendation result is unchanged even if the personalized sensitive features that the user cares about were changed in a counterfactual world. For example, if the user does not want to be discriminated by the recommender system on his or her age, then the system can guarantee that the recommendations (such as news articles) are age-neutral or age-diverse. \citeauthor{wu2022selective} \cite{wu2022selective} further advanced from personalized fairness to personalized selective fairness in recommendation, which provides users with the flexibility to select the sensitive feature(s) that they care about after the recommendation model is trained, and thus users can freely re-select the sensitive feature(s) that they care about when they want.

\subsubsection{\textbf{Blackbox Fairness vs. Explainable Fairness}}
Most fairness in recommendation research focus on defining fairness and developing methods to improve the fairness. However, an even more fundamental problem is to understand why a model is unfair.

\begin{itemize}
   \item
    \textbf{Blackbox Fairness}: Blackbox fairness models aim to develop methods to enhance fairness without trying to explain or understand the reasons of unfairness. 
   \item
   \textbf{Explainable Fairness}: Explainable fairness models focus more on how to understand the reasons that lead to unfair model outputs. 
\end{itemize}

There have been research on explaining recommendation results \cite{zhang2020explainable,tan2021counterfactual,zhang2014explicit,chen2022measuring,geng2022recommendation,xian2021ex3,ji2023counterfactual,chen2023dark}, explaining graph neural networks \cite{tan2022learning,lucic2022cf,ying2019gnnexplainer,yuan2020xgnn,xian2019reinforcement}, explaining vision and language models \cite{li2021personalized,goyal2019counterfactual,hendricks2018grounding,costa2018automatic,li2020generate}, etc.,
but the research on explaining why a model is fair or unfair is still very limited.
Understanding the ``why'' is not only helpful on technical perspectives but also on social perspectives. Technically, knowing the reasons of unfairness helps system designers to conduct data curation so as to remove the factors that lead to unfairness and also helps them to develop targeted models for disparity mitigation. Socially, knowing the reasons of unfairness helps policy makers to understand the social causes and implications of unfairness and develop prevention policies for future improvements. Explainable fairness is especially important for recommender systems because recommendation algorithms usually work with a huge amount (thousands or even millions) of both latent and explicit features in a collaborative learning way. In many other intelligent decision making systems such as loan approval and school admission, the total number of features is much smaller than recommender systems, and most importantly, it is usually straightforward for policy makers to know which feature(s) are sensitive that should be excluded from decision making so as to guarantee fairness, such as gender and race. However, recommender systems are different, on one hand, the huge amount of features makes it difficult to manually identify the sensitive features that lead to unfairness; on the other hand, most features are not immediately related to commonly known sensitive features such as race and gender but they still implicitly influence the model fairness; finally, even if the recommendation algorithm does not use explicit features at all and only uses user behaviors for model training, it may still result in unfair recommendations due to collaborative learning since some users' preferences and choices will influence other users' received recommendations. As a result, explainable fairness methods are highly demanded in recommender systems that can help to detect and explain why a model is unfair and how to improve. 
\citeauthor{ge2022explainable} \cite{ge2022explainable} aim to develop explainable fairness models for recommendation. The authors take item exposure fairness as an example and develop a counterfactual explainable fairness framework to explain which item features in the system significantly influence the model fairness. By reducing the influence of the detected sensitive features, the model is able to achieve better fairness-utility trade-off in recommendation.

\subsubsection{\textbf{Centralized Fairness vs. Federated Fairness}}

Another taxonomy of fairness in recommendation is the \textit{Centralized Fairness} and \textit{Federated Fairness}.

\begin{itemize}
   \item
    \textbf{Centralized Fairness}: Centralized fairness develops a central algorithm that has access to all users' data when trying to improve the fairness of the system.
   \item
   \textbf{Federated Fairness}: Federated fairness develops a decentralized algorithm that improves the fairness of the system without accessing all of the users' data, instead, each user's data is kept on the user's local device.
\end{itemize}

Recommendation accuracy is largely related to the level of details about the collected user information.
As a typical example, in the user cold-start recommendation scenario where the system has no previous records of the new user, the recommender system would benefit from transferring user knowledge from other sources to make a more accurate prediction.
However, this raises privacy concerns \cite{jeckmans2013privacy,bilge2013survey} for users and has motivated the recent research on federated recommender systems (FedRec) \cite{yang2020federated,fierimonte2016fully,ammad2019federated,han2021deeprec,anelli2021put,lin2020fedrec}.
Specifically, federated recommendation introduces the Federated Learning (FL) scheme \cite{mcmahan2017communication} into the optimization of recommender systems.
User data is maintained on the edge devices without being uploaded to the cloud server and thus guarantees user privacy.
The federated learning framework trains the recommendation model in a distributive manner with user edge devices as clients.
However, the federated learning paradigm also brings new challenges for fairness in recommendation.
On one hand, some existing fairness objectives in recommendation, especially user-based fairness objectives, may have an intrinsic conflict with horizontal federated learning \cite{zhou2021towards,liu2022fairness} -- 
while the evaluation and optimization of user fairness usually involve the collection of sensitive user features (e.g. gender, age, and sexual orientation), the federated learning framework wants to protect this information.
To overcome this conflict, one may have to relax or transform the fairness objective or design a privacy protection module for the user feature, to accommodate the federated recommender system \cite{liu2022fairness}.
On the other hand, recent research works have also pointed out that the federated system itself may introduce the collaboration fairness issue \cite{zhou2021towards,liu2022contribution} which emphasizes the equalized contribution between participants.
This fair federated learning problem has been studied in some federated learning research fields such as healthcare, computer vision, and natural language processing, and \citeauthor{zhou2021towards} \cite{zhou2021towards} provided a survey on this emerging topic.
In contrast, the research on fairness of federated recommender systems is still very limited \cite{liu2022fairness}.

\subsection{Methods}

Several methods have been proposed to mitigate unfairness in recommendation. The research on fair recommendation usually first define the fairness metrics they concern and then develop suitable techniques to promote the corresponding metrics. The relevant techniques are diverse and can be very different under different research and fairness definitions.
However, we try to organize them into several categories in this section and show some typical and common-used methods in each category so as to help readers better understand how fairness is technically achieved.

\subsubsection{\textbf{Regularization and Constrained Optimization}}

So far, the techniques for promoting fairness in recommendation are mainly in the form of regularization and constrained optimization. The various fairness criteria can be formulated as regularizers or constraints to guide the process of model optimization \cite{yao2017beyond, yao2017new, xiao2017fairness, burke2017balanced, farnadi2018fairness, zhu2018fairness, beutel2019fairness, li2021user}. The objective can be maximizing the utility of recommendation under fairness constraints \cite{yao2017beyond}, or maximizing fairness requirements under utility bounds \cite{zhu2018fairness}, or jointly optimizing both fairness and utility goals with a reasonable trade-off \cite{gao2019fair}.
Existing works usually apply regularization or constraint optimization to in-processing methods \cite{yao2017beyond, beutel2019fairness} and post-processing methods \cite{geyik2019fairness, li2021user}. The challenges for regularization and constrained optimization methods are that they are often non-convex in nature and are difficult to balance the conflicting constraints which may result in unstable training \cite{caton2020fairness}. A very clear example is \cite{yao2017beyond}. The authors propose four new unfairness metrics for preference prediction in collaborative filtering based recommendation, all measuring the discrepancy on prediction quality between the disadvantaged users and advantaged users. The disadvantaged and advantaged users are divided based on a binary group feature such as gender. Let's use $\mathrm{E}_{g}[r]_{j}$ and $\mathrm{E}_{\neg g}[r]_{j}$ to denote the average ratings for the $j$-th item from the disadvantaged and advantaged users, respectively. $\mathrm{E}_{g}[y]_{j}$ and $\mathrm{E}_{\neg g}[y]_{j}$ denote the average predicted score for the $j$-th item from disadvantaged users and advantaged users, respectively. We show the first fairness metric \textit{Value Unfairness} here as an example, and it measures the inconsistency in signed estimation error across user groups as:
$$
U_{\text {val }}=\frac{1}{n} \sum_{j=1}^{n}\left|\left(\mathrm{E}_{g}[y]_{j}-\mathrm{E}_{g}[r]_{j}\right)-\left(\mathrm{E}_{\neg g}[y]_{j}-\mathrm{E}_{\neg g}[r]_{j}\right)\right|
$$




To promote fairness in recommendation, the learning objective of recommendation task is extended with a smoothed variation of the fairness metrics. For example, the outer absolute value is replaced with the squared difference.
All the four fairness metrics have straightforward subgradients and can be optimized by various subgradient optimization techniques. The authors solve the problem for a local minimum as follows:
$$
\min _{\theta} L_{Rec}(\theta)+U .
$$

Other examples include \cite{beutel2019fairness}, where the authors propose to measure fairness based on pairwise comparisons and offer a regularizer to encourage improving this metric during model training and thus improve fairness in the resulting rankings; In \cite{xiao2017fairness}, the authors study unfairness issue in group recommendation, where items are recommended for a group of users whose preferences can be different from each other. Several fairness metrics are proposed for group recommendation scenario. The problem is formulated as a multiple objective optimization problem with the fairness metric as a regularizer and is solved from the perspective of Pareto Efficiency.

\subsubsection{\textbf{Adversary Learning}}

Another typical technique to promote fairness is to take advantage of adversary learning \cite{edwards2015censoring, xie2017controllable, beutel2017data, elazar2018adversarial,  wang2019balanced,bose2019compositional,arduini2020adversarial,du2020fairness, beigi2020privacy}. The basic idea of mitigating unfairness through adversary learning is to learn fair representations through a min-max game between the main task predictor and an adversarial classifier. The predictor aims to learn informative representations for the recommendation task, while the goal of the adversarial classifier is to minimize the predictor’s ability to predict the sensitive features from the representations, and thus the information about sensitive features are removed from the representations to mitigate discrimination. For example, let's consider $\mathcal{L}_{Rec}$ to denote the loss of the recommendation task such as the pair-wise ranking loss \cite{rendle2012bpr} or the mean square error loss \cite{koren2009matrix}. $\mathcal{L}_A$ denotes the loss of the adversarial classifier for sensitive feature $S$ which can be a cross-entropy loss for implementation. The adversary learning loss can be as follows:
\begin{equation}\label{eq}
\mathcal{L}=\sum_{(u, v) \in \mathcal{D}}\Big(\mathcal{L}_{\text {Rec}}(u,v, y_{uv})
-\lambda \cdot \mathcal{L}_A \left(u, S\right)\Big)
\end{equation}
where the adversarial coefficient $\lambda$ controls the trade-off between recommendation performance and fairness. An advantage of adversary approaches is that they often treat the prediction model as a black-box so as to offer the model-agnostic flexibility. However, the adversary approaches are usually difficult to train and lack stability \cite{feng2019learning, beutel2019putting}. \citeauthor{beigi2020privacy} \cite{beigi2020privacy} build a recommendation model with attribute protection to counter private-attribute inference attacks in recommendation. The authors solve the problem through adversarial learning with two main components: the private attribute inference attacker and the Bayesian personalized recommender. The goal of the attacker is to infer user private-attribute information while the recommender aims to learn users' interests under the regularization of the attacker. \citeauthor{wu2021fairrec} \cite{wu2021fairrec} propose a fairness-aware approach based on decomposed adversarial learning for news recommendation to mitigate the unfairness brought by the biases of user sensitive features. The proposed method decomposes user embedding into two parts: a bias-aware user embedding to capture the bias information on sensitive features and a bias-free user embedding for making fairness-aware news ranking. Adversary learning is used for removing the bias information of sensitive features from user embeddings.

\subsubsection{\textbf{Reinforcement Learning}}

Some researches believe that recommendation is not only a prediction problem but a sequential decision problem, and suggest to model the recommendation problem as a Markov Decision Process (MDP) and solve the problem through Reinforcement Learning (RL) \cite{afsar2021reinforcement}. Some research on fairness in recommendation also follow the trend and consider RL methods to promote fairness through a long-term and dynamic perspective \cite{jabbari2017fairness, williams2019dynamic, wen2019fairness,liu2021balancing, zhang2021recommendation, wen2021algorithms, ge2021towards,huang2021achieving, ge2022toward}. In \cite{ge2021towards}, the authors consider the dynamic item exposure fairness in recommender systems where the item popularity will change over time with recommendation actions and user feedback. The problem is formulated as a Constrained Markov Decision Process by dynamically constraining the item exposure fairness metric at each iteration. In \cite{ge2022toward}, the authors study the problem of Pareto optimal fairness-utility trade-off in recommendation and propose a framework based on multi-objective reinforcement learning to learn a single parametric representation for optimal recommendation policies over the space of all possible preferences. In \cite{liu2021balancing}, the authors study the problem of interactive recommender systems. Considering that the user preferences and the system fairness status are constantly changing over time, the authors jointly represent the user preferences and the system's fairness status into the states of the MDP recommendation model through an RL framework to achieve long-term fairness dynamically. One challenge of leveraging RL to promote fairness is that there may be a risk of ``user tampering'' \cite{evans2021user}, which is a phenomenon that a recommender system may try to increase its long-term user engagement through manipulating the user's opinions, preferences and beliefs via its recommendations. The potential unfairness of users that different users are affected by different manipulation needs vigilance from researchers and system developers \cite{zhang2021recommendation}.

\subsubsection{\textbf{Causal Methods}}
Causal methods for promoting fairness have been widely studied in classification tasks \cite{kusner2017counterfactual, galhotra2017fairness, chiappa2018causal, nabi2018fair, glymour2019measuring, wu2019counterfactual,kilbertus2020fair} and have also recently been applied to the field of recommendation. The key objective of causal methods is to investigate the relationships underlying the data and model, including the causal effects between sensitive variables and decisions, as well as the dependency between sensitive and non-sensitive variables. For example, \citeauthor{wu2018discrimination} \cite{wu2018discrimination} study causal-based anti-discrimination method in ranking problem. The authors build a causal graph to identify and remove both direct and indirect discrimination in ranked data and reconstruct a fair ranking if discrimination is detected from the causal graph; \citeauthor{zheng2021disentangling} \cite{zheng2021disentangling} leverage causal inference to solve popularity bias in recommendation. The authors assume that the user click behaviour depends on both their interest and the item popularity. To mitigate the effect of popularity bias, the authors propose a general framework to disentangle user interest and popularity bias through learning two types of embeddings: interest embedding and popularity embedding. The final recommendations are generated by only the interest embeddings so that the popularity bias is removed; \citeauthor{huang2021achieving} \cite{huang2021achieving} study how to achieve counterfactual fairness for users in online recommendation through incorporating causal inference into bandits. The authors adopt soft intervention to model the arm selection strategy and use the d-separation set identified from the causal graph to develop a fair UCB algorithm, which promotes fairness through choosing arms that satisfy the counterfactual fairness constraint. Another commonly-used causal-based methods is re-weighting the instances based on Inverse-Propensity-Scoring (IPS) techniques \cite{rosenbaum1983central}, which are usually adopted to solve the biases in recommendation such as popularity bias \cite{yang2018unbiased} and selection bias \cite{schnabel2016recommendations}. The IPS methods assume that the biases are caused by the fact that treatments are not being randomly assigned, and thus they use the inverse propensity of assigning the treatment as the sample weights to remove the bias. The IPS methods are usually convenient to implement, but the score estimator may not properly handle large shifts in observational probability \cite{bonner2018causal} and the methods are usually specifically designed for a particular problem setting. Besides improving fairness, the causal methods can also provide model transparency in terms of how decisions are made \cite{henelius2014peek}. A recent work \cite{ge2022explainable} studies the problem of how to explain which feature(s) lead to item exposure unfairness in recommendation through counterfactual reasoning. Causal methods also have challenges in practice, for example,
the causal-based notions are usually defined based on intervention and counterfactual which are non-observable quantities and cannot always be computed from observational data \cite{makhlouf2020survey}; and causal methods usually require the prior knowledge of causal graphs or causal assumptions which may not always be accessible in practice \cite{salimi2019interventional}.

\subsubsection{\textbf{Other Methods}}

Several other techniques are used to promote fairness of recommendations. Examples include 1) data augmentation methods such as \cite{rastegarpanah2019fighting}, which proposes a strategy to improve the socially relevant properties such as individual or group fairness of a recommender system through adding antidote data. The proposed framework is developed from an existing pre-trained matrix-factorization-based recommender system, and provides the flexibility of not having to modify the original input data or recommendation algorithm. The authors consider to add the ratings of new users to the input which are chosen based on the corresponding measure so as to improve the fairness of recommender systems; 2) methods based on Variational Autoencoders (VAEs) such as \cite{borges2019enhancing}, which proposes to incorporate randomness into the operation of VAEs so as to mitigate the position bias in multiple rounds of recommendation. The authors introduce four different noise distributions and find that adding noise to the process of sampling values from VAE's latent representation can provide long term fairness for recommendation with an acceptable trade-off between fairness and recommendation quality such as NDCG; 3) methods based on self-distillation \cite{wu2022big}, which leverages the model predictions on the original data as a teacher to regularize the predictions on the augmented data with randomly dropped user behaviors. The authors observe that the training process of big recommendation models can result in unfair recommendation performance for cold users, and show that the proposed  self-distillation method can push the model to fairly capture the interest distributions of heavy and cold users. More techniques are used to improve fairness in other fields of machine learning \cite{caton2020fairness, wan2021modeling}, but their application in the recommendation scenario is very limited.

\subsection{Datasets and Settings for Fairness in Recommendation Research}\label{data}

\subsubsection{\textbf{Research Datasets for Fairness in Recommendation}}
In this section, we collect some datasets for fairness research in recommendation. The requirements for datasets in fairness research usually highly depend on the specific fairness definitions. For example, for fairness research on user side, researchers usually need to consider fairness definitions about users' sensitive features; while for fairness research on item side, datasets with item features such as item category may be preferred. Some works even do not consider the fairness on user/item sensitive features but instead on user activity or item popularity, which can be directly computed from the user interaction data, and thus they have no requirement for user/item features from the datasets. 
Compared with the datasets that contain user-item interactions, recommendation datasets that contain user/item sensitive features are very limited and uneasy to find. Therefore, we introduce several datasets for recommendation where the user/item sensitive feature information is available. All datasets are publicly available under the referenced links. The data statistics are shown in Table.\ref{data}.

\begin{table}[ht]
\centering
\small
\caption{Statistics of datasets. \textit{varies} means the dataset contains multiple sub-datasets (\textit{Amazon}) or multiple types of interactions (\textit{Alibaba}, \textit{XING-Rec 17}), so that the number may vary with different sub-datasets or prediction tasks.}
\begin{tabular}{lllll}
\hline
Dataset & Sensitive Feature & \#Users & \#Items & \#Interactions \\
\hline
MovieLens & gender, age and occupation & 6K & 3.7K & 1M \\
Last.FM & gender, age and country & 120K & 50M & 2B \\
Last.FM-360K & gender, age and country & 359K & 294K & 17M \\
RentTheRunWay & age & 105K & 5.8K & 192K \\
Electronics & gender & 1M & 9K & 1M \\
Alibaba & age and gender & 49M & 200M & varies \\
Insurance & gender, marital status and occupation &  1K &  21 & 5K \\
Post & gender & 500 & 6K & 72K \\
Coat & gender and age & 290 & 300 & 11K \\
Sushi & gender and age & 5K & 100 & 50K \\
Amazon & item category & varies & varies & varies \\
XING-Rec 17 & job title, membership and location, etc & 1.4M & 1.3M & varies \\
Yelp & food genre & 2.1M & 160K & 8.6M \\
ModCloth & body shape and product size & 45K & 1K & 100K \\
\hline
\end{tabular}
\label{data}
\end{table}

\begin{itemize}
\item {\textbf{MovieLens}.\protect\footnote{\url{https://grouplens.org/datasets/movielens/1m/}}} 
This is a dataset for movie recommendation tasks with 1,000,209 anonymous user-movie ratings. It has approximately 3,706 movies and 6,040 users with each rating within 1 to 5. It also has a smaller version dataset with 100,000 ratings, 943 users and 1682 movies. There are three user sensitive features: gender, age and occupation.

\item {\textbf{Last.FM}.\protect\footnote{\url{http://www.cp.jku.at/datasets/LFM-2b/}}} 
This is a large dataset for music retrieval and recommendation tasks~\cite{melchiorre2021investigating}. It has more than 2 billion user-music listening events, 120,322 users, 5,159,580 artists and 50,813,373 tracks. There are three user sensitive features: gender, age and country.

\item {\textbf{Last.FM- 360K}.\protect\footnote{\url{http://ocelma.net/MusicRecommendationDataset/lastfm-360K.html}}} 
This is a smaller music recommendation dataset collected from the Last.fm website~\cite{celma2009music}. It has 17 million records with 359,347 users and 294,015 artists. There are three user sensitive features: gender, age and country.

\item {\textbf{RentTheRunWay}.\protect\footnote{\url{https://cseweb.ucsd.edu/~jmcauley/datasets.html\#clothing_fit}}} This dataset contains user-cloth renting interactions~\cite{misra2018decomposing}. It has 192,544 interactions, 105,508 users and 5,850 items. User age is included as a sensitive feature. 

\item {\textbf{Electronics}.\protect\footnote{\url{https://cseweb.ucsd.edu/~jmcauley/datasets.html\#market_bias}}} This dataset contains user online shopping behaviors on Amazon for e-commerce recommendation. It has 1,292,954 shopping events, 1,157,633 users and 9,560 products. User gender information is included as a sensitive feature. Note that this gender feature is from predictions rather than  from user profile data. Details can be found in~\cite{wan2020addressing}.

\item {\textbf{Alibaba}.\protect\footnote{\url{https://github.com/rec-agent/rec-rl}}} 
This datast is from the Alibaba e-commerce platform~\cite{pei2019value} which contains user-page interactions. It considers a recommendation scenario where a user is recommended with 50 items per page in a ranked order. There are 49 million users and 200 million items. The interaction types include requests (refresh recommendation lists), clicks, adding to carts, adding to wishlists and purchases. Age and gender information are included as sensitive features.


\item {\textbf{Insurance}.\protect\footnote{\url{https://www.kaggle.com/mrmorj/insurance-recommendation}}} This is a Kaggle dataset with the goal of recommending insurance products to a target user. It has 5,382 interactions, 1,231 users and 21 insurance policies. There are three user sensitive features: gender, marital status and occupation.

\item {\textbf{Post}.\protect\footnote{\url{https://www.kaggle.com/datasets/vatsalparsaniya/post-pecommendation}}} This dataset is for post recommendation tasks. It has 71,800 user-post interactions, 500 users and 6,000 posts with user gender information as the sensitive feature.

\item {\textbf{Coat}.\protect\footnote{\url{https://www.cs.cornell.edu/~schnabts/mnar/}}}
This dataset is collected by \citeauthor{schnabel2016recommendations} \cite{schnabel2016recommendations}. They hire 290 Mechanical Turkers to select a set of items among 300 candidates and ask them to place ratings on selected items. Each user comes with 24 rated items in the training set and 16 items in the testing set. There are two sensitive features: gender and age.

\item {\textbf{Sushi}.\protect\footnote{\url{https://www.kamishima.net/sushi/}}}
This dataset is collected by \citeauthor{kamishima2003nantonac}~\cite{kamishima2003nantonac} which includes responses of a questionnaire survey of preference in SUSHI. It can be used for rating prediction or ranking tasks. There are 5,000 users and 100 SUSHI candidates. This dataset contains a ranking data which asks users to rank the order of SUSHI based on their personal preferences and a scoring data which asked users to give 1 to 5 rating to the ordered food. There are two sensitive features: gender and age.  

\item 
{\textbf{Amazon}.\protect\footnote{\url{https://cseweb.ucsd.edu/~jmcauley/datasets/amazon_v2/}}}
This dataset contains user and item profiles and interactions from Amazon \cite{ni2019justifying}. The website includes many sub-datasets that contain user and item information of various categories in Amazon such as the Amazon Beauty, Books, CDs and Vinyl, etc, and the data sizes vary. Although the sensitive features from user side are limited in these datasets, item-side fairness can be considered based on the available category information of items.

\item 
{\textbf{XING-Rec 17}.\protect\footnote{\url{http://2017.recsyschallenge.com/\#dataset/}}}
This dataset is released by XING, a career-oriented social networking site, to support the ACM RecSys Challenge 2017 for a job recommendation task \cite{abel2017recsys}. The dataset contains 1.4M users, 1.3M items, 310M impressions, 6.9M clicks, 282K bookmarks, 118K replies, 907K deletes and 101K recruiter interests for the offline setting. The dataset contains the user profiles such as job title, membership types, education degree, and location, as well as item profiles such as job title, required skills, and location, etc.

\item 
{\textbf{Yelp}.\protect\footnote{\url{https://www.yelp.com/dataset}}}
This datasets contains the user reviews for restaurants. The dataset contains 2.1M users, 160.5K businesses, and 8.6M interactions. Existing research works usually use some subsets of the original dataset based on their own pre-processing methods \cite{serbos2017fairness, zhu2020measuring, ge2022explainable}. The food genres can be considered as a sensitive feature for items such as ``American,'' ``Japanese,'' ``Italian,'' and ``Chinese'' \cite{zhu2020measuring}.

\item 
{\textbf{ModCloth}.\protect\footnote{\url{https://github.com/MengtingWan/marketBias}}}
This dataset is collected from an e-commerce website selling women's clothing and accessories \cite{wan2020addressing}. The dataset contains 100K reviews from 45K users and 1K items. The body shapes of users, as well as the product sizes of items can be considered as the sensitive features.

\end{itemize}

\subsubsection{\textbf{Research Settings for Fairness in Recommendation}}

Besides the datasets, we also introduce the basic experimental settings for evaluating the fair recommendation methods. It is worth noting that the experimental datasets, data pre-processing strategies, and evaluation metrics for fair recommendations vary a lot depending on the specific fairness definitions, proposed methods and application domains. In this section, we try to provide a general overview of these components commonly used in the evaluation of fair recommendations.

For the basic experimental settings, researches usually consider several datasets containing user profiles, item profiles and user-item interactions (e.g., ratings, clicks, views). Based on specific fairness considerations, the dataset may need to contain the sensitive feature information from user or item sides. For the data pre-processing strategies, researches usually need to handle the missing data through dropping the N/A or estimating the missing values based on other available data. For fairness works, the feature engineering process needs to take the fairness considerations into account, for example, works may need to avoid the use of sensitive features as input features to achieve the \textit{No Disparate Treatment} \cite{barocas2016big}, which seeks equality of treatment by prohibiting the use of the sensitive features in the decision process.

For evaluation metrics, fair recommendation methods usually need to show their effectiveness on both the accuracy metrics and fairness metrics. For accuracy metrics, the common evaluation metrics for recommendation include NDCG, Hit Rate, Precision, Recall, and F1-score, which measure the accuracy of the recommended items compared to the ground truth. However, the fairness metrics are closely related to what kind of fairness definitions are considered in the work, for example, the group-level fairness can be measured by equalized odds \cite{berk2021fairness}, equal opportunity \cite{hardt2016equality}, or statistical parity \cite{dwork2012fairness} difference, while the individual-level fairness can be measured by the treatment difference between similar individuals \cite{dwork2012fairness}. Overall, the choice of experimental settings and evaluation metrics may depend on the specific fairness goals and recommendation tasks. Usually researchers would adapt the settings and evaluation methods specific to their fair recommendation approaches and datasets.

\section{Challenges and Opportunities} \label{challenges}

So far, researchers have realized the importance of improving fairness in recommender systems and have started the exploration. However, the research in this area are still very limited and many important problems have not even been studied. In this section, we talk about some open challenges and point out some future opportunities with the hope to inspire more research works in this area.

\subsection{Lack of Consensus on Fairness Definitions}

The most challenging fact in fairness research is that there is no unique consensus on fairness definition. As we mentioned above, in essence, fairness in machine learning requires that "equal individuals/groups should be treated equally". However, how to determine "the equal individuals/groups" and what form of "equality treatment" is acceptable depends on the certain circumstances. The fairness considerations are different in different scenarios, the biases that lead to unfairness are diverse, and the fairness demands can be put forward from several different perspectives. The lack of consensus on fairness definition leads to a series of problems.

First, how to achieve multiple fairness requirements at the same time.
Existing methods are usually designed for achieving one particular type of fairness requirement. However, this is not sufficient for solving the problem since people's demand for fairness is diverse and various biases often occur simultaneously \cite{chen2020bias}. 
As a result, it is imperative to explore whether it is possible to propose a unified model to handle multiple fairness demands. 
Though a single solution to all fairness problems is impossible since it has been theoretically proven that some fairness notions are inherently conflicting with each other and cannot be achieved at the same time \cite{kleinberg2016inherent}, it is still worth it to explore some simple scenarios such as handling two or three different cases. Furthermore, it is interesting to explore the relationship between various biases and fairness notions and clarify whether we really need to address so many fairness definitions. It could be possible that as long as the system guarantees several important fairness notions then most other fairness notions will be naturally guaranteed or guaranteed to some extent, and thus we do not have to work on indefinitely many fairness notions.
Second, when some fairness requirements cannot be achieved simultaneously due to their conflicts, one important problem is how to achieve a good and reasonable balance. 
Solving the problem may not only need technical innovations such as Pareto optimality but also careful social and ethical considerations such as transparency and trust.
Third, it would be important to explore personalized fairness which allows each individual user to express his or her own fairness demand \cite{li2021counterfactual}, so that the system would not have to resolve the conflicting fairness definitions just based on the system designers' understanding.
Finally, if the system is designed for achieving only one fairness requirement, it is important to know how to choose the most suitable fairness consideration for the specific case. For example, which one is more important for improving user fairness in recommendation? Group fairness or individual fairness? Static fairness or dynamic fairness? To answer the questions, it is important to study how to evaluate different fairness notions for a specific scenario. This further leads to another challenge: since existing works are usually proposed under different perspectives, they often adopt different or even unique evaluation metrics to evaluate fairness. Therefore, it is sometimes difficult to compare the existing methods. To this end, some evaluation metrics or benchmark datasets that can help us to compare different fairness notions from a unified view are highly appreciated in the future.


\subsection{Relationship between Fairness and other Trustworthy AI Perspectives}

Fairness is a very important perspective for trustworthy, responsible and ethical AI research. However, except for fairness, there are many other perspectives to consider, such as explainability, controllability, robustness and privacy. A very important problem is to explore the relationship between fairness and other trustworthy AI perspectives and how can the different perspectives interact with each other to develop better trustworthy AI systems. 

Take fairness and explainability as an example. As mentioned above, due to the inherent conflict between some fairness definitions, in some cases, the system may not be able to meet every user's fairness requirement due to the conflicting fairness demands, and in such cases, transparency and honesty from the system may be highly appreciated. For instance, the system may honestly explain to some users why they have to bear a bit of unfair treatment at the current moment in order to better serve other (usually the majority of) people or for certain long-term goals, and how the system will compensate such unfair treatment in the future. Besides, if we implement a fair system, we also need to explain to users why a decision is fair for the user so as to make sure users trust the intelligent system. This requires the research community to better understand the relationship between explainability and fairness and develop explainable fairness \cite{ge2022explainable} models in the future so that explainability and fairness can benefit each other. Other research topics include the relationship between fairness and robustness \cite{ovaisi2022rgrecsys} such as out-of-distribution generalization, the relationship between fairness and privacy such as the fairness in federated learning \cite{liu2022fairness}, and the relationship between fairness and controllability such as if and how users can actively control the fairness demands for themselves \cite{li2021counterfactual,wu2022selective}.

\subsection{Causal Foundations for Fairness}
As we introduced in section \ref{causal}, it is important to consider causal-based fairness notions to address unfairness in recommendation. There are some works considering causality in classification \cite{li2021general,kusner2017counterfactual} and general recommendation \cite{xu2021causal,xu2021deconfounded,li2022causal,xu2022dynamic,xu2023causal} tasks, however, the exploration of causal fairness in recommendation problem is still very limited. Although recommendation task can be formulated as a classification problem in some cases, the causal fairness techniques in classification may not be directly migrated to the recommendation problem. For example, to achieve counterfactual fairness in classification, as stated in \cite{kusner2017counterfactual}, the most straightforward way to guarantee the independence between the predicted results and the sensitive features is just to avoid using the sensitive features (and the features causally depend on the sensitive features) as input. However, this is not the case in recommendation scenarios: most of the collaborative learning-based recommendation algorithms such as collaborative filtering and collaborative reasoning are directly trained from user-item interaction information \cite{ekstrand2011collaborative} which do not use any feature, no matter sensitive feature or non-sensitive feature. However, the model still generates unfair recommendations on user sensitive features even if the model does not directly use any feature as input \cite{li2021counterfactual}. The reason is that during collaborative learning, the model may capture the underlying relationships between user features and user behaviours that are inherently encoded into the training data, since user features may have causal impacts on user behaviors and preferences. As a result, we need to design methods to achieve counterfactually fair recommendations and this cannot be realized in trivial ways such as not using sensitive feature, because the collaborative learning model does not use sensitive feature from the beginning. Therefore, more explorations about causality considerations are needed to understand the underlying reasons of unfairness and solve the unfairness issue properly.

\subsection{Understanding the Connection between Bias and Fairness}

Bias and fairness are often mentioned concurrently or even obfuscated sometimes, but their relationship has not been clearly understood or discussed in the community. In general, various biases could be the main causes of the unfair results in recommendation. For example, the popularity bias in data may cause exposure unfairness on item side, while the gender bias in data may cause unfair treatment on user side. The connection between a specific type of bias and unfairness can be intricate. For example, unfairness in gender may be caused by biases in the training data, biases in the model design, or the biases of using inappropriate optimization functions or evaluation benchmarks. Moreover, a particular bias, such as bias in the training data, can also result in multiple forms of unfairness, such as unfairness towards individuals, unfairness towards groups, short-term unfairness, or long-term unfairness. What's more, the presence of bias does not necessarily always lead to unfairness. For example, even if there is bias in the training data, unfairness can be avoided through model calibration. Besides the various biases in data or algorithm, there are also other reasons for unfairness. For example, researches have shown that some fairness requirements can not be satisfied at the same time \cite{chouldechova2017fair, kleinberg2016inherent, pleiss2017fairness}, therefore, the violation of one type of fairness may be caused by ensuring another. 

Recently, researchers have explored various biases and debias methods in recommendation \cite{chen2020bias}, though being obfuscated in some cases, the works about bias and unfairness in recommendation usually have some differences. Firstly, existing debias researches usually focus on how to use debias methods to improve the recommendation accuracy or robustness, rather than promoting fairness. Secondly, fairness works usually present clear fairness definitions and well-defined quantitative metrics to evaluate model unfairness, for example, papers may use the differences on model performance across groups to measure group-level unfairness. The works about fairness in recommendation usually evaluate the proposed methods on both the fairness metrics and the model accuracy metrics. However, the works about debias usually evaluate the model performance on metrics of model accuracy, and use the improvement on the accuracy to show the effectiveness of debiasing. What's more, though biases are the main causes of unfairness and the debias methods may be helpful to enhance fairness, many works on fairness are not implemented through debiasing methods but through directly adding fairness requirements over the model outcome, such as adding a fairness constraint on the optimization process, which bears the risk of losing accuracy. Therefore, there is a gap between the research on debias and fairness though the two problems have deep connections both theoretically and practically. As a result, the relationship between bias and unfairness should be carefully considered and the connection between debiasing and fairness promotion methods should be better established, which may lead to better understandings on the causes of unfairness as well as better methods for improving both fairness and accuracy.

\subsection{Fairness of Large Language Model based Recommendation}

With the recent development and success of Large Language Models (LLMs), LLM-based recommendation has become a new frontier of recommender system modeling and research, which enables Universal Recommendation Engine (URE) \cite{geng2022recommendation} and Artificial General Recommender (AGR) \cite{lin2023sparks,ge2023openagi}. For example, as a pioneer work, researchers proposed the P5 model \cite{geng2022recommendation,xu2023openp5,geng2023vip5,hua2023up5}, which is a Pre-train, Personalized Prompt and Predict Paradigm (P5) for recommendation. The model formulates recommender system as a language generation task and designs a set of prompts for each recommendation task, including rating prediction, sequential recommendation, direct recommendation, explanation generation, and review summarization, thus enabling a universal recommendation engine by formulating various recommendation tasks using one model, one loss, and one data format \cite{geng2022recommendation,hua2023index}. However, there are new challenges in terms of ensuring the fairness of LLM-based recommendation models. For example, even though the language prompt does not explicitly include sensitive feature words such as the exact gender and age of the user, some other words that are indirectly related to sensitive feature words may still lead the model to unfair recommendations \cite{li2021personalized}, such as ``pretty'' which may indirectly reveal the user's gender and ``young'' which may indirectly reveal the user's age. There have been recent attempts to explore and address the fairness issues of large language models. For example, \citeauthor{li2023fairness} \cite{li2023fairness} explored the fairness of ChatGPT on fundamental classification tasks in high-takes fields such as education, criminology, finance and healthcare, and found that LLM such as ChatGPT is more fair than small models, though there are still some unfairness issues. \citeauthor{zhang2023chatgpt} \cite{zhang2023chatgpt} conducted an evaluation of ChatGPT and found that it still exhibits unfairness to some sensitive attributes when generating recommendations. Furthermore, \citeauthor{hua2023up5} \cite{hua2023up5} went beyond just evaluating the fairness of LLM-based recommendation and instead aim to improve the fairness of LLM-based recommendation models. More specifically, they developed an unbiased foundation model (UP5) for fairness-aware recommendation, which is based on the Counterfactually-Fair-Prompting (CFP) technique. Though some prelimiary attempts have been made on the fairness of LLM-based recommendation, more efforts are needed to make LLM-based recommender systems more fair, unbiased, and reliable.

\subsection{Considerations in Real-World Deployment}

Although many fairness research have been conducted in academia and industry, deploying fairness-aware models to benefit users in real-world system requires a lot more practical considerations.
The deployment of fairness-aware models in industry can be broadly classified into two types: user-oriented deployment and developer-oriented deployment. User-oriented deployment focus on delivering fairness-aware results to real users of the system and thus directly influence the service of users, while developer-oriented deployment do not immediately deliver such results to users but instead focus on developing tools within the production environment to help developers and policy makers better understand the system unfairness. It is worth noting that the ``users'' in user-oriented deployment not only include the average users of the system such as the consumers in e-commerce or the viewers in short video social networks, other stakeholders in the system such as the various sellers in e-commerce and the video creators in social networks are also users of the platform, who also have their fairness demands such as fair exposure opportunity of their items.

An example of user-oriented deployment is LinkedIn \cite{geyik2019fairness}, which explored the first large-scale deployed framework for ensuring fairness in the hiring domain. Researchers applied the fairness framework to LinkedIn Talent Search so as to achieve fairness criteria such as equality of opportunity and demographic parity in candidate ranking. Researchers also presented the online A/B testing results, which showed that their approach resulted in tremendous improvement in fairness metrics without affecting the business metrics.
An example of developer-oriented deployment is Amazon SageMaker Clarify \cite{hardt2021amazon}, which developed explainability toolkits for Amazon SageMaker and was deployed in the Amazon AWS cloud computing clusters. Based on the toolkits, various developers over the world who use the cloud cluster can easily detect and monitor if there is any bias in their data or model and if the results produced by their model are fair. This can help the developers to better understand the consequences of their models and thus help them to better refine the models.
Due to the multi-stakeholder nature of fairness in real-world systems \cite{abdollahpouri2019multi}, in the future, it is important to understand the difference and relationship between the various fairness demands from various stakeholders in real-world systems, and how the knowledge from developer-oriented deployment can be transferred to user-oriented deployment so as to directly benefit the users.

\subsection{Understanding the Relationship between Fairness and Utility}
As mentioned above, fairness-aware methods may be deployed in user-oriented or developer-oriented ways in practice.
Developer-oriented deployment usually meet fewer obstacles in real-world systems because it does not immediately influence the users and business metrics. However, more in-depth considerations or trade-offs usually need to be considered for user-oriented deployment in practical systems.
On one hand, real-word recommender systems are usually very complex with multiple modules and massive training data, and thus it can be difficult to figure out all of the fairness problems in such a huge system. On the other hand, many industry companies usually put business metrics such as purchase rates first compared to the fair treatment of users, and thus the incentive to promote fairness is less than that of chasing profits in many practical systems, especially if there is an inevitable trade-off between fairness and profit metrics. 
Of course, one approach to promoting fairness in practice is through various legal requirements that are already implemented in many countries, such as the General Data Protection Regulation (GDPR) in EU\footnote{\url{https://gdpr.eu/}}, California Consumer Privacy Act (CCPA) in the US\footnote{\url{https://oag.ca.gov/privacy/ccpa}}, and the Internet Information Service and Algorithmic Recommendation Regulation (IISARR) in China\footnote{\url{http://www.cac.gov.cn/2022-01/04/c_1642894606364259.htm}}, which enforce fair treatment of users in real-world systems. However, on the other hand, it is also very important for the research community to further explore the relationship between fairness and utility so as to improve the incentive of industry practitioners to promote fairness. 

This requires further explorations on three folds: 1) Explore the cases where fairness and utility do not conflict with each other but actually promote each other. Actually, there have been research on user-oriented group-fairness in recommendation \cite{li2021user,rahmani2022experiments} which show that by carefully defining the groups and fairness metrics, it is possible for recommendation models to improve both fairness and utility (such as recommendation accuracy) at the same time. In the future, it is promising to explore if co-improving fairness and utility is possible under other fairness and utility definitions;
2) If there is really an inevitable conflict between certain fairness and utility metrics, we can develop methods to guarantee fairness with as few utility losses as possible or guarantee utility with as few fairness losses as possible, explain such losses to users in appropriate ways so that they can accept it, and amortize the losses across users or time so that the system does not always hurt some (types of) users. This may require further innovations on multi-objective optimization, Pareto optimization, explainable fairness, dynamic fairness and creative user interfaces in the future;
3) Same as any other trustworthy AI research, understanding the benefits of fairness in recommendation should be considered in the dynamic and long-term context. Some research show that certain fairness and utility metrics may have to trade-off with each other in short-term settings \cite{ge2022toward,wang2021understanding}, however, due to the limitations on datasets and evaluation methods, existing research on fairness-utility relationship are mostly conducted under static environments using fixed datasets rather than in dynamic deployed systems. As a result, it is not surprising to observe that increasing fairness may hurt recommendation accuracy because recommending more long-tail items to increase exposure fairness will naturally lead to fewer recommendation hits in static datasets where popularity bias of user clicks are already recorded in the testing set. This does not necessarily mean that increasing fairness definitely hurts utility in deployed systems, because in the long-term, if users feel they are fairly treated by the system, it will help to increase users' retention, interest, trust and engagement in the system, and will thus help to create a sustainable eco-system in the platform, which eventually leads to increased utility and profits for the platform in the long-run. As a result, it is important to explore the relationship between utility and fairness in the long-run in real-world deployed systems, and this requires close collaboration between academia and industry in the future.

\subsection{Developing Better Metrics, Benchmarks and Simulation Environments}

Though we have introduced several benchmark datasets to study fairness in this survey, the available dataset to support fairness research is still very limited. This is because the research on fairness usually requires datasets that have sufficient features such as user personal information. Besides, there exist many different fairness concerns and each requires certain dataset to study, which means that a diverse portfolio of datasets are needed for fairness research. As a result, extensive efforts are needed in the future to create various datasets to support fairness study. More importantly, this may require innovative solutions to meet the demands of both researchers and data providers. For example, many fairness research rely on users' personal information, however, most of the personal information are users' sensitive features such as gender, race, age, location and income that are closely related to users' privacy. As a result, researchers should take extreme care when creating and publishing datasets for fairness research such as proper anonymization of users and even developing advanced methods for both privacy protection and fairness promotion, such as through federated learning or differential privacy. 

On the other hand, many users in real-world systems may choose to not reveal their personal information in registration or even provide wrong personal information. The reason for users to do so is exactly because they are afraid of being unfairly treated by the system or breach of their privacy. However, without such information, it would be difficult to improve fairness for them even if the system want to do so, which results in a dilemma due to the crisis of confidence. As a result, it is important to create a good incentive and build trust with users so that they are willing to provide the relevant information to support fairness promotion. Besides, the community can explore reliable data augmentation and data curation methods to improve the quality of incomplete datasets, and develop fairness-aware methods that do not rely on users' sensitive features when they are not available.

Fairness evaluation is also an important problem that needs further exploration. As mentioned in previous sections, the fairness definitions are very diverse. Currently, each type of fairness has its unique evaluation methods, which makes it difficult to compare different fairness methods or develop unified models that can improve fairness from multiple perspectives. As a result, it would be promising if a systematic evaluation protocol can be developed which enables us to evaluate diffident fairness requirements on the same scale. Furthermore, we may even develop methods to evaluate various utility and trustworthy considerations in the same framework, such as fairness, accuracy, explainability, robustness and privacy. Finally, to support the evaluation of dynamic fairness and to explore the long-term effect of fairness, a simulation environment will be highly appreciated by the community. We are pleased to see that few recent works such as \cite{rahmani2022experiments, boratto2022consumer} began to explore comparing different fairness methods in recommendation. However, these works only cover one perspective of fairness in recommendation, i.e., the user fairness, since it is hard to measure so many different aspects of fairness within a same framework. Researchers have attempted to build simulation platforms for recommendation accuracy research such as RecSim \cite{ie2019recsim}, and similar platforms for fairness research will be very useful not only to the recommender system community but also to the broader fairness in AI communities.





\section{Conclusions}\label{con}

In this survey, we aim to introduce the foundations, definitions, methods and outlooks for fairness in recommendation research. We begin the survey with a brief introduction of fairness in machine learning to provide beginners with a general view and basic background knowledge of fairness research, including the causes, methods, as well as different considerations of fairness. To help better understand the fairness concepts in recommendation, we also introduce the basic fairness definitions and methods in classification and ranking tasks since they are closely related to fairness in recommender systems. For the main body of the survey, we provide a taxonomy to classify the existing fairness definitions in recommender systems to help reader build a systematic understanding of the area. We also introduce the various technical and evaluation methods as well datasets for fairness in recommendation research. 
Finally, we discuss the challenges and opportunities of fairness research in recommendation with the hope of both inspiring future innovations and promoting fairness deployment in real-world systems.





\bibliographystyle{ACM-Reference-Format}
\bibliography{main}


\end{document}